\documentclass[twocolumn,showpacs,preprintnumbers,amsmath,amssymb,pra]{revtex4}
\usepackage{graphicx}
\usepackage{epsf,dsfont,hyperref,url,float}

\def\figsubcap#1{\par\noindent\centering\footnotesize(#1)}

\begin{document}

\title{Quantum Game Theory and Open Access Publishing}

\author{Matthias Hanauske}
\author{Steffen Bernius}
\affiliation{
Johann Wolfgang Goethe-University\\
Institute of Information Systems\\
Mertonstr. 17, 60054 Frankfurt$/$Main
}

\author{Berndt Dugall}
\affiliation{
Johann Christian Senckenberg University-Library\\
Bockenheimer Landstr. 134-138, 60325 Frankfurt$/$Main
}

\date{\today}

\begin{abstract}  
The digital revolution of the information age and in particular the sweeping changes of scientific communication brought about by computing and novel communication technology, potentiate global, high grade scientific information for free. The arXiv for example is the leading scientific communication platform, mainly for mathematics and physics, where everyone in the world has free access on. While in some scientific disciplines the open access way is successfully realized, other disciplines (e.g. humanities and social sciences) dwell on the traditional path, even though many scientists belonging to these communities approve the open access principle. In this paper we try to explain these different  publication patterns by using a game theoretical approach. Based on the assumption, that the main goal of scientists is the maximization of their reputation, we model different possible game settings, namely a zero sum game, the prisoners' dilemma case and a version of the stag hunt game, that show the dilemma of scientists belonging to ``non-open access communities''. From an individual perspective, they have no incentive to deviate from the Nash Equilibrium of traditional publishing. By extending the model using the quantum game theory approach it can be shown, that if the strength of entanglement exceeds a certain value, the scientists will overcome the dilemma and terminate to publish only traditionally in all three settings. 
\end{abstract}

\pacs{03.67.-a, 02.50.Le, 01.20.+x, 01.30.Xx, 89.65.-s, 89.70.+c}
\maketitle

\section{Introduction}\label{sec:Rep}
In recent years the market of scientific publishing faces several forces that may cause a major change of traditional market mechanisms. First of all, the increase of digitalization brought a shift towards electronic publication. Furthermore, shrinking library budgets with a simultaneous rise of journal prices resulted in massive cancellations of journals and books \cite{getz-1999,tenopir_2000,wellcome_2003,mccabe-2004}. In consequence of this still lasting “journal crisis”, alternative ways of publishing, in particular open access, received increasing attention \cite{dugall-2004,okerson-1996,tananbaum-2003}. Currently two main approaches have emerged. On the one hand, new open access journals are brought to being, either through transformation of traditional journals or through creation of new titles. This approach is often called the “Golden Road to Open Access”. On the other hand, authors may self-archive their articles in Institutional Repositories, a model referred to as the “Green Road to Open Access” \cite{harnad-2005,guedon-2004}. 

The realization of open access publishing differs between research disciplines \cite{eu_2006}. The prime example of an adoption of the open access publishing paradigm is the arXiv server which is mainly used by physicists and mathematicians. Researchers in this fields normally self-archive their papers on the arXiv (so that everyone has free access to the work) and often additionally submit them to regular scientific journals, where these papers go through the traditional peer review process. Thus the arXiv-model represents neither exactly the golden nor the green road of open access publishing. 

In contrast most other scientific disciplines do not make use of open access publishing, even though they support this model if asked for \cite{dfg_2003,schroter-2005}. Instead, they submit research papers to traditional journals that do not provide free access to their articles. Considering that the majority of scientists regard open access publishing as superior to the traditional system, the question arises, why it is only adopted by few disciplines. 

Based on the assumption, that the main goal of scientists is the maximization of their reputation, we try to answer this question from the perspective of the producers of scientific information by using a game theoretical approach. Scientific reputation originates mainly from two different sources: on the one hand the citations to the articles of a scientist and on the other hand the reputation of the journals he publishes his articles \cite{dewett-2004}. Starting from a general 2-Scientists-Game, where two authors have to decide whether they publish open access or not, three different possible game settings are developed. In each case the outcome of the game results in a dilemma, that cannot be solved within the static framework of classical game theory. Therefore we extend the model using the quantum game theoretical approach and show, that if choosing quantum strategies, the players can escape the dilemma. 

The remainder of the paper is structured as follows. In section \ref{sec2} the open access game is developed using the classical game theoretical notation. Firstly we define the general reputation payoff matrix of the game. The three settings of the game cover a zero sum game, the prisoners' dilemma case, and a variation of the so called stag hunt game. In section \ref{sec3}, after a brief introduction into the history of quantum game theory, we define the basic notations of the quantum version of the open access game and discuss the different game settings in detail. Our results are summarized in section \ref{sec4}. 

\section{The Classical Game of Open Access}\label{sec2}
\subsection{Formalization of the Game}
\begin{figure}[H]
\vspace*{-1.2cm}
\centerline{
\includegraphics[width=4.3in]{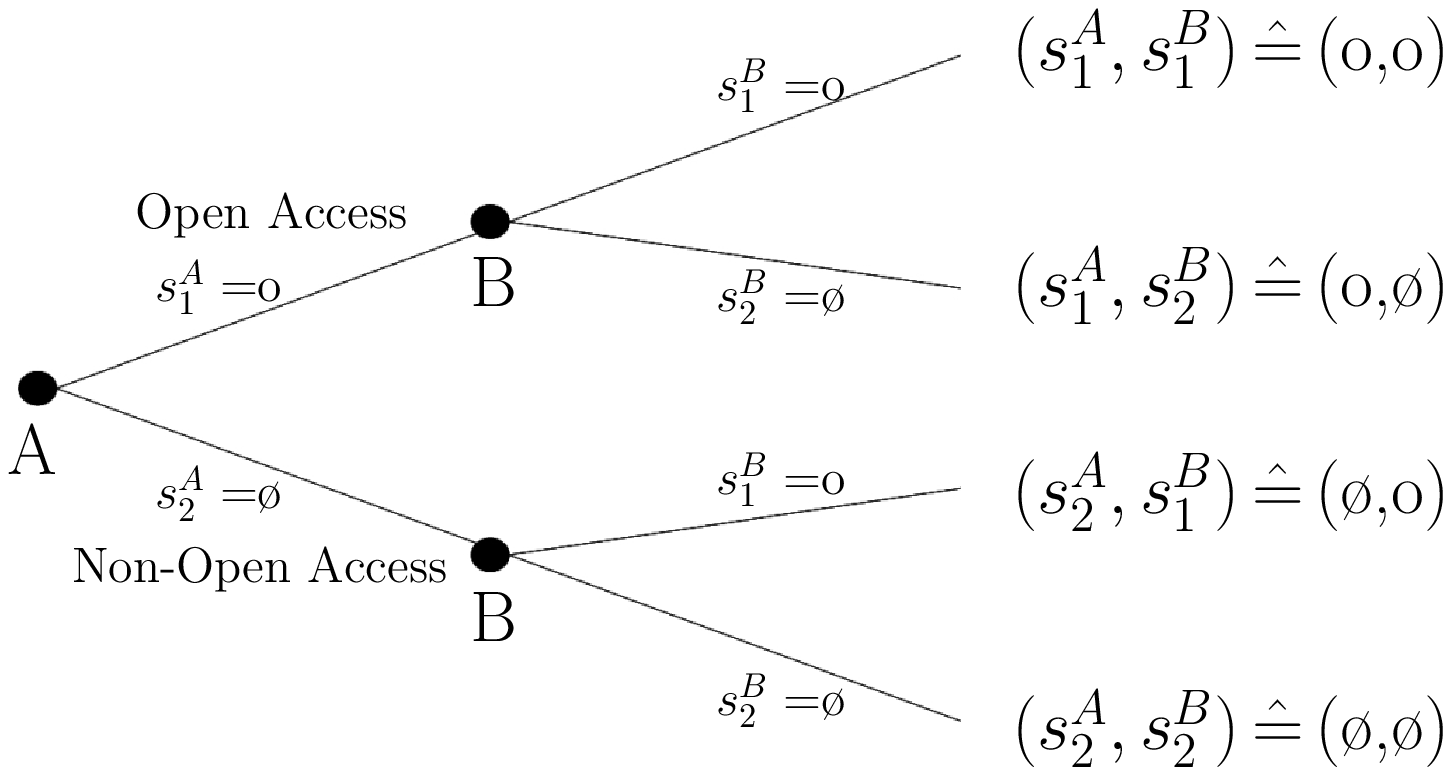}
}
\vspace*{-10.2cm}
\caption{Classical tree of the open access game.}
\label{fig:clgametree}
\end{figure}

To describe the classical open access game we use a normal-form representation of a two-player \footnote{In reality, the open access game consists of a lot of players. One can therefore understand Player B moreover as an overall construct of the probabilistic choice of the whole scientific community in which A is embedded.} game $\Gamma$ where each player (Player 1 $\hat{=}$ A, Player 2 $\hat{=}$ B) can choose between two strategies (${\cal S}^{A}=\{s^{A}_1,s^{A}_2\}$, ${\cal S}^{B}=\{s^{B}_1,s^{B}_2\}$). In our case the two strategies represent the authors' choice between publishing open access (o) or not (\o). The game tree can therefore be visualized as in Fig. \ref{fig:clgametree}.
The whole strategy space ${\cal S}$ is composed with use of a Cartesian product of the individual strategies of the two players (scientists): 
\begin{equation}
{\cal S} =  {\cal S}^A \times {\cal S}^B = \left\{ \hbox{(o,o)}, \hbox{(o,\o)}, \hbox{(\o,o)}, \hbox{(\o,\o)} \right\}
\end{equation}
As outlined in the introduction, we assume, that the main objective of scientists is the maximization of their reputation. In the following we focus on a situation, where the two scientists belong to a scientific community in which the open access paradigm is not yet broadly adopted and the publishers decline the acceptance of articles that are already accessible on an open access server. The payoff structure of this game can be described by the following matrix:
\begin{table}[H]
\centerline{
\begin{tabular}{|r|c c|}
        \hline
        A$\setminus$B& o & \o
        \\ \hline
        o & ($r+\delta$,$r+\delta$) & ($r-\alpha$,$r+\beta$)
        \\
        \o & ($r+\beta$,$r-\alpha$) & ($r$,$r$)\\\hline
\end{tabular}}
\caption[caption]{General open access payoff matrix.}
\label{tab:PayOff_general}
\end{table}
The actual reputation of the two scientists is represented by a single parameter $r$ \footnote{By using this formalization, we assume that both scientists are on a similar level of reputation. It can be shown that if they have different ``starting'' reputation values,  the outcome of the classical game would be the same.}. If both players decide to publish their papers only in  traditional journals (\o,\o), their reputation $r$ does not change. If only one of the two players chooses the open access strategy ((\o,o) or (o,\o)) the parameters $\alpha$ and $\beta$ ($\alpha,\beta\geq 0$) describe the decrease and the increase of the scientists' reputation, depending on the selected strategy. By modeling the payoff in this way, it is assumed that the reputation of the player, who performs open access, decreases if the other player simultaneously decides not to publish open access. This can be explained by the fact, that in ``non-open access communities'' reputation is mainly defined through the reputation of the journals a scientist publishes in. Thus if performing open access (by what a publication in traditional journals gets impossible), the scientist has no chance to gain journal-related reputation any more. On the other hand the parameter $\beta$ describes the potential increase of reputation of a scientist that refuses to perform open access while the other player selects the open access strategy. By setting $\alpha = \beta$ the reputation is considered as a relative construct (see section \ref{sec:RelQ}). The parameter $\delta$ represents the potential benefit in the case that both players choose the open access strategy (o,o). The payoff for each player then is $r+\delta$. In this case it is assumed that if all players choose the open access strategy the publishers are forced to accept articles for publication even if they are already accessible. Then scientists can gain reputation both through the reputation of the journal they publish in and through the increase of citations due to a broader accessibility \cite{lawrence-2001,harnad-2004,eysenbach-2006}. 

In the following we will describe three specific parameter settings of the open access game.   
\subsection{Potential Game Settings}
\subsubsection{Open Access as a Zero Sum Game}\label{sec:RelQ}
The most simple case of an open access game is realized by setting the free parameters of the games' payoff matrix to the following fixed values: $r=0$, $\delta=0$ and $\alpha=\beta=1$. The starting reputation and the open access benefit of both players is set to zero, whereas the absolute value of the increase ($\beta$) and decrease ($\alpha$) in reputation is taken to be equal. This setting therefore describes reputation as a relative quantity. A potential increase in reputation of one player results in an equivalent decrease of the other player's reputation. In this case, $\delta$ has to be zero because the total amount of reputation in the system cannot increase. The payoff matrix of this setting is illustrated in Table \ref{tab:PayOff_case1}.
\begin{table}[H]
\fboxrule0.7mm
\centerline{
\begin{tabular}{|r|c c|}
        \hline
        A$\setminus$B& o & \o
        \\ \hline
        o & ($0$,$0$) & ($-1$,$1$)
        \\
        \o & ($1$,$-1$) & \fbox{($0$,$0$)}\\\hline
\end{tabular}}
\caption[caption]{Open access payoff matrix with reputation as a relative quantity.}
\label{tab:PayOff_case1}
\end{table}
In this game each player has a dominant strategy (\o) and the Nash equilibrium is (\o,\o). Therefore no player has the incentive to deviate from the non-open access strategy \o. 

\subsubsection{The Open Access Game as a Prisoners' Dilemma}
The game is similar to a classical prisoners' dilemma, if the assumption that reputation is a relative quantity is partially abrogated. If both players choose the open access strategy, the total amount of reputation will increase by $\delta$ ($\delta > 0$). In this case we have taken the following parameter settings: $r=3$, $\delta=1$ and $\alpha=\beta=2$. Table \ref{tab:PayOff_case2} depicts the payoff of both players.
\begin{table}[H]
\fboxrule0.7mm
\centerline{
\begin{tabular}{|r|c c|}
        \hline
        A$\setminus$B& o & \o
        \\ \hline
        o & ($4$,$4$) & ($1$,$5$)
        \\
        \o & ($5$,$1$) & \fbox{($3$,$3$)}\\\hline
\end{tabular}}
\caption[caption]{Open access payoff matrix within the prisoners' dilemma setting.}
\label{tab:PayOff_case2}
\end{table}
Although the payoff for both players would be higher if they choose the strategy set (o,o), they are stuck within the Nash equilibrium (\o,\o). This outcome describes the paradox situation of many scientific disciplines: Scientists on the one hand realize that they would benefit, if all players adopt open access, but on the other hand, no player has an individual incentive to change. 

\subsubsection{Open Access as a ``Stag Hunt'' Game}
The stag hunt game in its original meaning describes the situation of two hunters, which have the choice between hunting a stag or a rabbit. If successful, bagging a stag provides more benefit than bagging a rabbit. The problem within this game is that hunting a stag can only be successful if both players go for the stag, whereas a rabbit can be easily bagged by only one hunter. In our case hunting a stag corresponds to the strategy of performing open access, and the non-open access strategy stands for hunting rabbits. Compared to the prisoners' dilemma only the parameter $\beta$ is modified. To formulate the open access stag hunt game we have used the following parameter settings: $r=3$, $\delta=1$, $\alpha=2$ and $\beta=0$ (see Table \ref{tab:PayOff_case3}) \footnote{In contrast to the original stag hunt game, where hunting a stag alone results in a payoff of zero, in this case the single open access performer gets a payoff of 1, simply because a reputation value of zero is unrealistic. A reputation value of zero only makes sense, if reputation is seen as a relative quantity (see section \ref{sec:RelQ}).}.
\begin{table}[H]
\fboxrule0.7mm
\centerline{
\begin{tabular}{|r|c c|}
        \hline
        A$\setminus$B& o & \o
        \\ \hline
        o & \fbox{($4$,$4$)} & ($1$,$3$)
        \\
        \o & ($3$,$1$) & \fbox{($3$,$3$)}\\\hline
\end{tabular}}
\caption[caption]{Open access payoff matrix within the stag hunt setting.}
\label{tab:PayOff_case3}
\end{table}
In contrast to the other settings this game has two pure Nash equilibria ((o,o) and (\o,\o)) and one mixed strategy Nash equilibrium $\frac{2}{3}$(o,o). (o,o) is payoff dominant, whereas (\o,\o) is the risk dominant pure Nash equilibrium. The mixed strategy Nash equilibrium $\frac{2}{3}$(o,o) implies that one scientist has the incentive to choose non-open access if he expects the probability of the other player to choose non-open access as well, to be higher than $33.\overline{3}\,$\%. 

In the following section we formulate the classical game settings described above within a  quantum game theoretical framework. 

\section{The Quantum Game of Open Access}\label{sec3}
The basic principles of game theory were developed by J. von Neumann in the year 1928. Together with O. Morgenstern he applied this new theory to economics \cite{book_neumann_gametheory}. In addition to this outstanding scientific contribution he was also involved in the description of the mathematical foundations of quantum theory \cite{book_neumann_quantumtheory}. Keeping these historical facts in mind, it is surprising, that only recently game theory and quantum physics has been unified to one theory, the so called {\it Quantum Game Theory}.

The leadoff articles of quantum game theory where published by D. A. Meyer and J. Eisert et al. in the year 1999. Meyer illustrated a quantum version of the simple ``Penny Flip'' game and showed, that if one player uses a specific quantum strategy, whereas the other player persists in a classical one, the player who selects the quantum strategy will always win the game \cite{meyer-1999-82}. Just a few weeks after Meyers' article was published, Eisert et al. focused on the well known prisoners' dilemma \cite{eisert-1999-83}, unknowing Meyers' results. Within their quantum representation they where able to demonstrate, that prisoners could escape from the dilemma, if the entanglement of the prisoners' wave function is above a certain value. S. C. Benjamin and P. M. Hayden amplified the formal description of quantum games towards many players \cite{benjamin-2001-64}. L. Marinatto and T. Weber applied the density matrix approach to the ``Battle of Sexes'' game and demonstrated, that entangled strategies lead to a unique solution of the game \cite{marinatto-2000-272}. E. W. Piotrowski and J. Sladkowski disposed quantum game theory to market behaviors \cite{piotrowski-2002-312}. In 2001 J. Du et al. realized the first simulation of a quantum game; the experimental results confirmed their theoretical predictions \cite{du-2002-88}.  Particularly they performed a prisoners' dilemma quantum game on their nuclear magnetic resonance quantum computer. Several other topics regarding quantum game theory have been addressed (e.g. overviews are given in \cite{flitney-2002-2,grabbe-2005,iqbal-2006}).

In the following subsection we summarize the main formal concepts of a two-player two-strategy quantum game. We follow the description of Eisert et al. \cite{eisert-1999-83,eisert-2000-47} and allow two parameter sets of quantum strategies \footnote{This limitation of allowed quantum operations corresponds to the allowed set $S^{(TP)}$ in \cite{eisert-2000-47}.}.

\subsection{Formalization of the Quantum Game}\label{sec:quantumgame}
One can understand the concept of quantum strategies as an enlargement of mixed strategies towards an abstract complex strategy space. The measurable classical strategies (o and \o) correspond to the orthonormal unit basis vectors $\left| \hbox{o} \right>$ and $\left| \hbox{\o} \right>$ of the two dimensional complex space $\mathds{C}^2$, the so called Hilbert space ${\cal{H}}_i$ of the player $i$ ($i=A,B$). A quantum strategy of a player i is represented as a general unit vector $\left| \psi \right>_i$ in his strategic Hilbert space ${\cal{H}}_i$. The whole quantum strategy space $\cal{H}$ is constructed with the use of the direct tensor product of the individual Hilbert spaces: ${\cal{H}}:={\cal{H}}_A \otimes {\cal{H}}_B$. The main difference between classical and quantum game theory is, that in the Hilbert space ${\cal{H}}$ correlations between the players' individual quantum strategies are allowed, if the two quantum strategies $\left| \psi \right>_A$ and $\left| \psi \right>_B$ are entangled. The overall state of the system we are looking at is described as a two-players quantum state $\left| \Psi \right> \in {\cal{H}}$. The four basis vectors of the Hilbert space ${\cal{H}}$ are chosen to be equal to the classical game outcomes ($\left| \hbox{oo} \right>$, $\left| \hbox{o\o} \right>$, $\left| \hbox{\o{o}} \right>$ and $\left| \hbox{\o\o} \right>$). 

The setup of the quantum game begins with the choice of the initial state $\left| \Psi_0 \right>$. We assume that both players are in the state $\left| \hbox{o} \right>$. The initial state of the two players is then given by $\left| \Psi_0 \right> \,=\, \hat{\cal{J}}  \left| \hbox{oo} \right>$, where the unitary operator $\hat{\cal{J}}$ is responsible for the possible entanglement of the two player system. The players' quantum decision (quantum strategy) is formulated with the use of a two parameter set of unitary $2\times2$ matrices: 
\begin{eqnarray}
&\hat{{\cal{U}}}(\theta,\varphi) :=
\left(
\begin{array}[c]{cc}
e^{i\,\varphi} \, \hbox{cos}(\frac{\theta}{2})&\hbox{sin}(\frac{\theta}{2})\\
-\hbox{sin}(\frac{\theta}{2})&e^{-i\,\varphi} \, \hbox{cos}(\frac{\theta}{2})
\end{array}
\right)&\\
&
\forall \quad \theta \in{} [0,\pi] \,\, \wedge \,\, \varphi \in{} [0,\frac{\pi}{2}] &\quad .\nonumber
\end{eqnarray}
By arranging the parameters $\theta$ and $\varphi$ a player is choosing his quantum strategy. The classical strategy o for example is selected by appointing $\theta=0$ and $\varphi=0$ :
\begin{equation}
\hat{{\cal{\mbox{o}}}}:=
\hat{{\cal{U}}}(0,0) =
\left(
\begin{array}[c]{cc}
1&0\\
0&1
\end{array}
\right)\quad,
\end{equation}
whereas the strategy {\o} is selected by choosing $\theta=\pi$ and $\varphi=0$ :
\begin{equation}
\hat{{\cal{\mbox{\o}}}}:=
\hat{{\cal{U}}}(\pi,0) =
\left(
\begin{array}[c]{cc}
0&1\\
-1&0
\end{array}
\right)\quad.
\end{equation}

\begin{figure}[ht]
\vspace*{-0.8cm}
\centerline{
\includegraphics[width=3.8in]{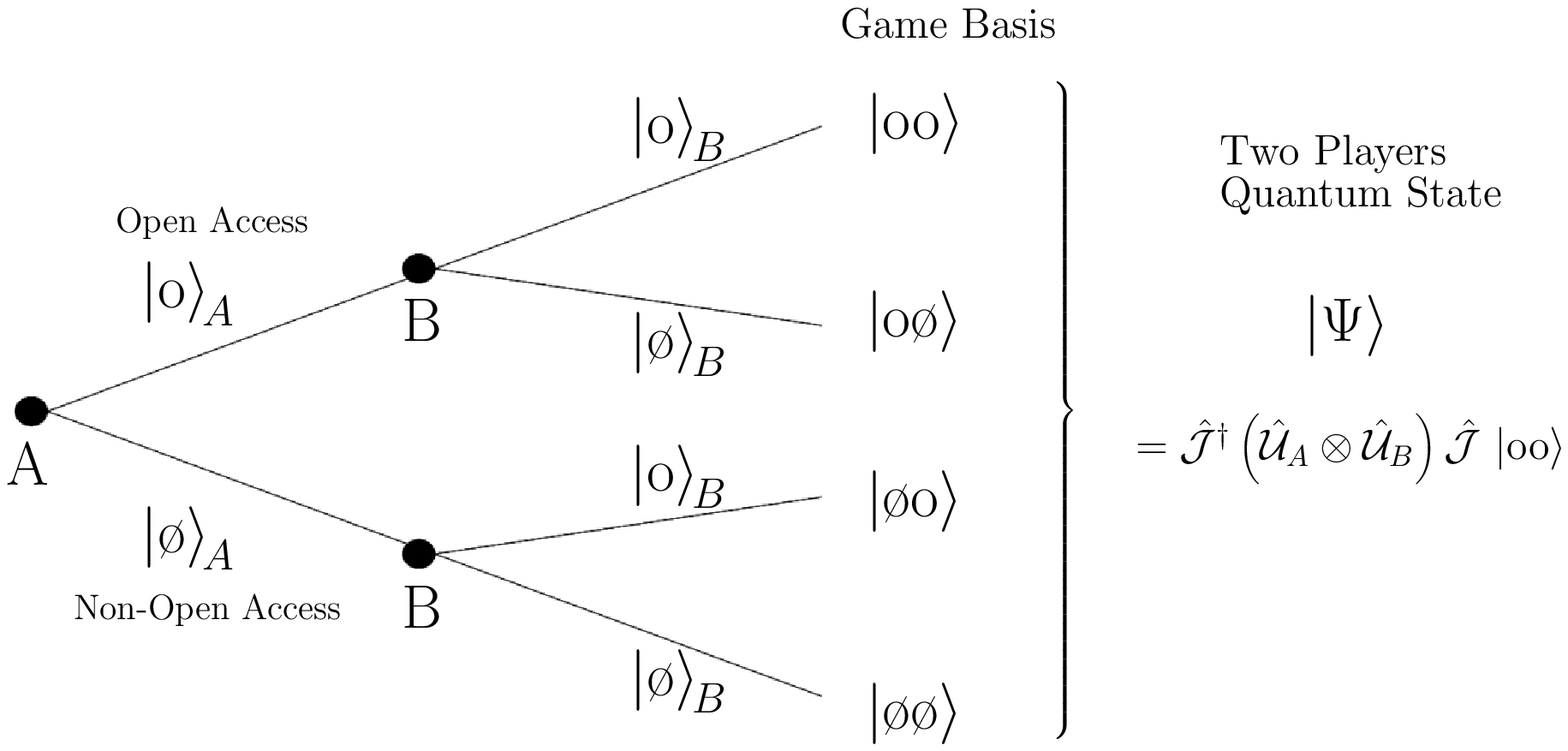}
}
\vspace*{-9.0cm}
\caption{Tree of the open access quantum game.}
\label{fig:quantumtree}
\end{figure}

The tree of the open access quantum game is displayed in Fig. \ref{fig:quantumtree}.
After the two players have chosen their individual quantum strategies ($\hat{{\cal{U}}}_A:=\hat{{\cal{U}}}(\theta_A,\varphi_A)$ and $\hat{{\cal{U}}}_B:=\hat{{\cal{U}}}(\theta_B,\varphi_B)$) the disentangling operator $\hat{\cal{J}}^\dagger$ is acting to prepare the measurement of the scientists' state. The entangling and disentangling operator ($\hat{\cal{J}}, \hat{\cal{J}}^\dagger$; with $\hat{\cal{J}}\equiv\hat{\cal{J}}^\dagger$) is depending on one additional single parameter $\gamma$ which is a measure of the entanglement of the system:
\begin{equation}
\hat{\cal{J}}:= e^{i \, \frac{\gamma}{2} (\hat{{\cal{\mbox{\o}}}} \, \otimes \, \hat{{\cal{\mbox{\o}}}})} \,\, , \quad \gamma \in{} [0,\frac{\pi}{2}] \quad .
\end{equation}
The final state prior to detection therefore can be formulated as follows:
\begin{equation}
\left| \Psi_f \right> = \hat{\cal{J}}^\dagger \left( \hat{\cal{U}}_A \otimes \hat{\cal{U}}_B \right) \hat{\cal{J}}\, \left| \mbox{oo} \right> \quad .
\end{equation}
The expected payoff of the two scientists within the quantum version of the open access game depends on the payoff matrix (see Table \ref{tab:PayOff_general}) and on the joint probability to observe the four possible outcomes of the game:  
\begin{eqnarray}
\$_A&=& (r+\delta)\,P_{\mbox{\small  oo}} + (r-\alpha)\,P_{\mbox{\small o\o}} + (r+\beta)\,P_{\mbox{\small \o{o}}} + r\,P_{\mbox{\small \o\o}} \nonumber\\
\$_B&=& (r+\delta)\,P_{\mbox{\small oo}} + (r+\beta)\,P_{\mbox{\small o\o}} + (r-\alpha)\,P_{\mbox{\small \o{o}}} + r\,P_{\mbox{\small \o\o}} \nonumber\\
&&\mbox{with:}\quad P_{\sigma \sigma^{,}}=\left| \, \left< \sigma\sigma^{,} | \Psi_f \right> \, \right|^2 \,\, , \quad \sigma,\sigma^{,}=\left\{\mbox{o},  \mbox{\o} \right\} \quad .\nonumber
\end{eqnarray}
To visualize the payoffs in a three dimensional diagram it is neccessary to reduce the set of parameters in the final state: $\left| \Psi_f \right>=\left| \Psi_f(\theta_A,\varphi_A,\theta_B,\varphi_B) \right> \rightarrow \left| \Psi_f(t_A,t_B) \right>$. We have used the same specific parameterization as Eisert et al. \cite{eisert-1999-83}, where the two strategy angles $\theta$ and $\varphi$ depend only on a single parameter $t \, \in{} [-1,1]$. In our model $t_A,t_B=1$ corresponds to strategy $\mbox{\o}$, and $t_A,t_B=0$ corresponds to strategy $\mbox{o}$. Negative $t$-values correspond to quantum strategies, where $\varphi>0$. 

\begin{figure}[ht]
\vspace*{0.4cm}
\centerline{
\includegraphics[width=4in]{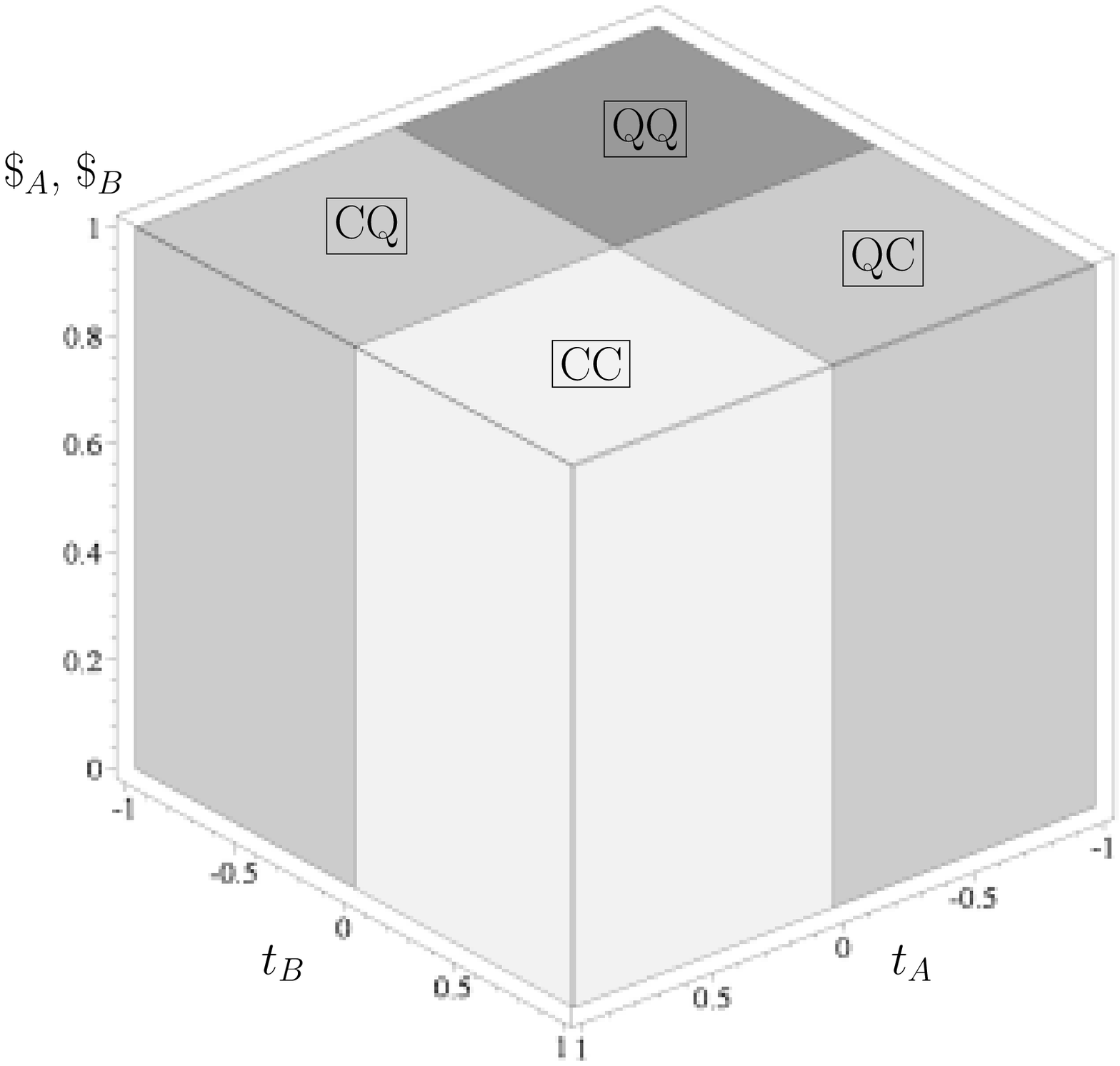}
}
\vspace*{-6.5cm}
\caption{Separation of the strategy space in four different regions; namely the absolute classical region CC, the absolute quantum region QQ, and the partially classical-quantum regions CQ and QC.}
\label{fig:stratreg}
\end{figure}

Fig. \ref{fig:stratreg} shows the general structure of the separation of strategy regions.
The whole strategy space is separated into four regions, namely the absolute classical region (CC: $t_A,t_B\geq0$), the absolute quantum region (QQ: $t_A,t_B<0$) and the two partially classical-quantum regions (CQ: $t_A\geq0 \wedge t_B<0$ and QC: $t_A<0 \wedge t_B\geq0$). In the following subsection we will present the main results of the different game settings of the open access quantum game. The outcomes of the different games are illustrated by visualizing the payoff surfaces of scientist A and scientist B as a function of their strategies $t_A$ and $t_B$. 

\subsection{Potential Game Settings}
\subsubsection{Open Access as a Zero Sum Quantum Game}
\begin{figure}[H]
\vspace*{0.35cm}
\centerline{
\includegraphics[width=4.0in]{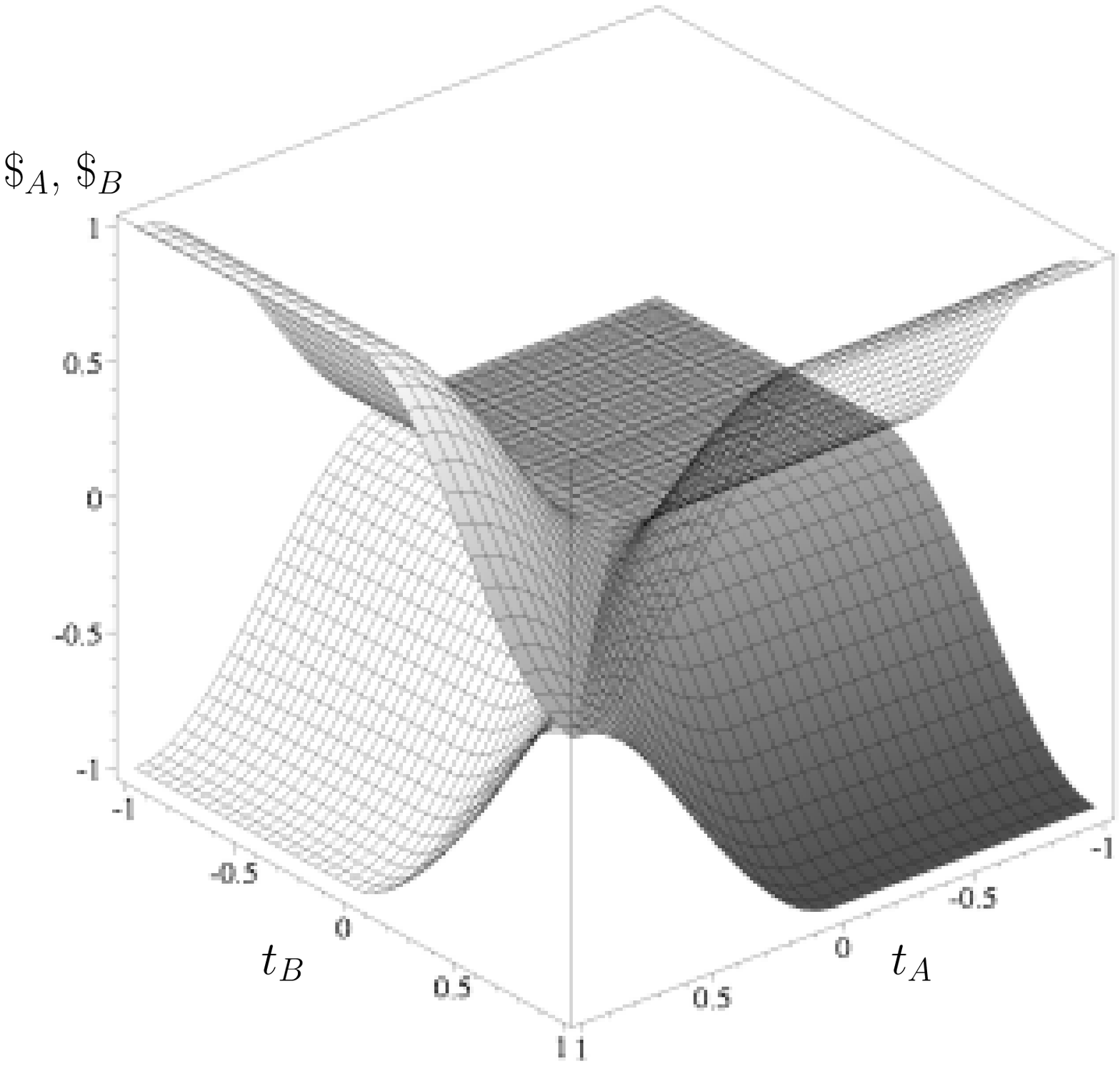}
}
\vspace*{-6.4cm}
\caption{Expected payoff of scientists A and B in a separable quantum game (payoff setting see Table \ref{tab:PayOff_case1}).}
\label{fig:1sep}
\end{figure}

Using the simple payoff matrix (Table \ref{tab:PayOff_case1}) and the quantum game formulation of section \ref{sec:quantumgame} we have calculated the expected payoff for the two scientists with and without entanglement. Fig. \ref{fig:1sep} depicts the expected payoff for scientist A ($\$_A$, intransparent surface) and scientist B ($\$_B$, wired surface) as a function of their strategies $t_A$ and $t_B$ in a separable quantum game 
($\gamma=0$). 
The outcome of this separable quantum game is similar to the classical solution outlined in section \ref{sec:RelQ}. To illustrate this, we recall the definitions of dominant strategies and Nash equilibria and formulate them in respect to our possible quantum choices: 

($\theta_A^\star,\varphi_A^\star;\theta_B^\star,\varphi_B^\star$) is a dominant quantum strategy if
\begin{eqnarray}
\$_A(\hat{{\cal{U}}}_A^\star,\hat{{\cal{U}}}_B) &\geq& 
\$_A(\hat{{\cal{U}}}_A,\hat{{\cal{U}}}_B)\qquad   
\forall \quad
\hat{{\cal{U}}}_A\,\, \wedge \,\,\hat{{\cal{U}}}_B \\
\$_B(\hat{{\cal{U}}}_A,\hat{{\cal{U}}}_B^\star) &\geq& 
\$_B(\hat{{\cal{U}}}_A,\hat{{\cal{U}}}_B) \qquad   
\forall \quad
\hat{{\cal{U}}}_A\,\, \wedge \,\,\hat{{\cal{U}}}_B \quad .
\nonumber
\end{eqnarray}

($\theta_A^\star,\varphi_A^\star;\theta_B^\star,\varphi_B^\star$) is a quantum Nash equilibrium if
\begin{eqnarray}
\$_A(\hat{{\cal{U}}}_A^\star,\hat{{\cal{U}}}_B^\star) &\geq& 
\$_A(\hat{{\cal{U}}}_A,\hat{{\cal{U}}}_B^\star) \qquad   
\forall \quad
\hat{{\cal{U}}}_A \\
\$_B(\hat{{\cal{U}}}_A^\star,\hat{{\cal{U}}}_B^\star) &\geq& 
\$_B(\hat{{\cal{U}}}_A^\star,\hat{{\cal{U}}}_B) \qquad   
\forall \quad
\hat{{\cal{U}}}_B \quad .
\nonumber
\end{eqnarray}

In the classical version of the game there exists one dominant strategy, namely (\o,\o), which corresponds to the parameter set ($\theta_A^\star=\pi,\varphi_A^\star=0 \, , \, \theta_B^\star=\pi,\varphi_B^\star=0$). The expected payoff in this dominant strategy is equal to zero for both players ($\$_A(1,1)=\$_B(1,1)=0$, see Fig. \ref{fig:1sep}). Because of the validity of the following conditions, (\o,\o) is also a dominant strategy in the separable game: 
\begin{eqnarray}
&\$_A(t_A=1,\hat{{\cal{U}}}_B) = {\mbox{cos}\left(\frac{\theta_B}{2}\right)}^2  \quad \geq\quad \$_A(\hat{{\cal{U}}}_A,\hat{{\cal{U}}}_B) = &\nonumber\\ 
 &= {\mbox{sin}\left(\frac{\theta_A}{2}\right)}^2 {\mbox{cos}\left(\frac{\theta_B}{2}\right)}^2  - 
{\mbox{cos}\left(\frac{\theta_A}{2}\right)}^2 {\mbox{sin}\left(\frac{\theta_B}{2}\right)}^2 \label{eq:sep1a} \quad ,&\\
&\$_B(\hat{{\cal{U}}}_A,t_B=1) = {\mbox{cos}\left(\frac{\theta_A}{2}\right)}^2  \quad \geq\quad \$_B(\hat{{\cal{U}}}_A,\hat{{\cal{U}}}_B) = &\nonumber\\ 
&= {\mbox{sin}\left(\frac{\theta_B}{2}\right)}^2 {\mbox{cos}\left(\frac{\theta_A}{2}\right)}^2  - 
{\mbox{cos}\left(\frac{\theta_B}{2}\right)}^2 {\mbox{sin}\left(\frac{\theta_A}{2}\right)}^2 \label{eq:sep1b}\quad .&
\end{eqnarray}

The conditions (\ref{eq:sep1a}) and (\ref{eq:sep1b}) are easy to illustrate if one examines Fig. \ref{fig:1sep}. To visualize condition (\ref{eq:sep1a}) for example, one shall look at the intransparent surface and fix an arbitrary point on the surface, which is located on the curve $\$_A(1,t_B)$ (with $t_B \in [-1,1]$). Condition (\ref{eq:sep1a}) means, that if one varies $t_A$ between all possible strategies ($t_A \in [-1,1]$), while keeping $t_B$ fixed, the payoff of player A ($\$_A$) will always decrease. In a similar way, condition (\ref{eq:sep1b}) can be illustrated by considering the wired surface $\$_B(t_A,t_B)$. 

Recapitulating the separable zero sum open access quantum game, one can say that no changes to the classical game are observable. Due to the dominance of strategy (\o,\o), both scientists will not perform open access. 

The situation is entirely different in the maximally entangled version of the game.
\begin{figure*}
\vspace*{0.45cm}
\begin{center}
  \parbox{3.3in}{\includegraphics[width=3.2in]{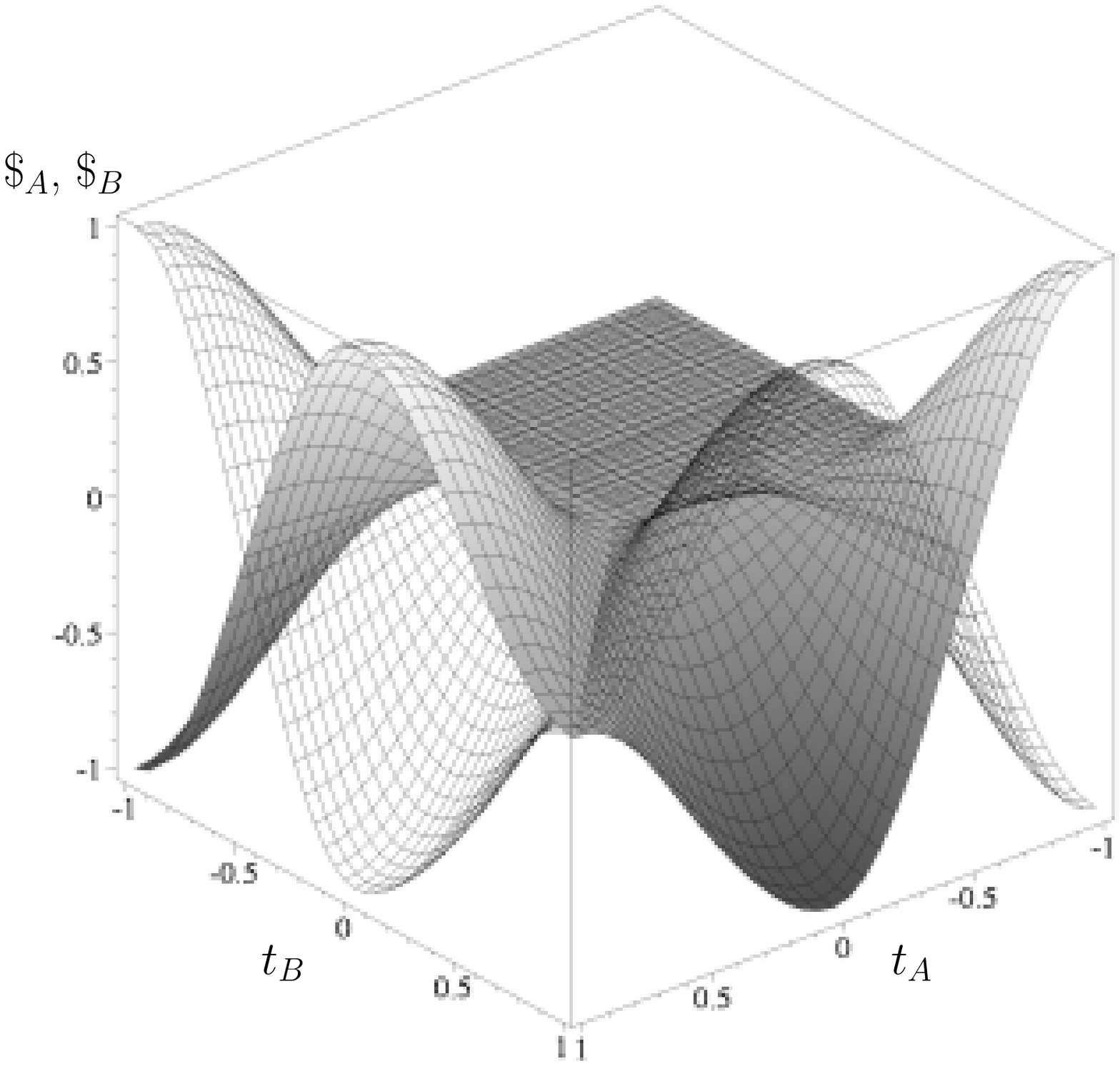}\vspace*{-4.7cm}\figsubcap{a}}
  \parbox{3.3in}{\includegraphics[width=3.2in]{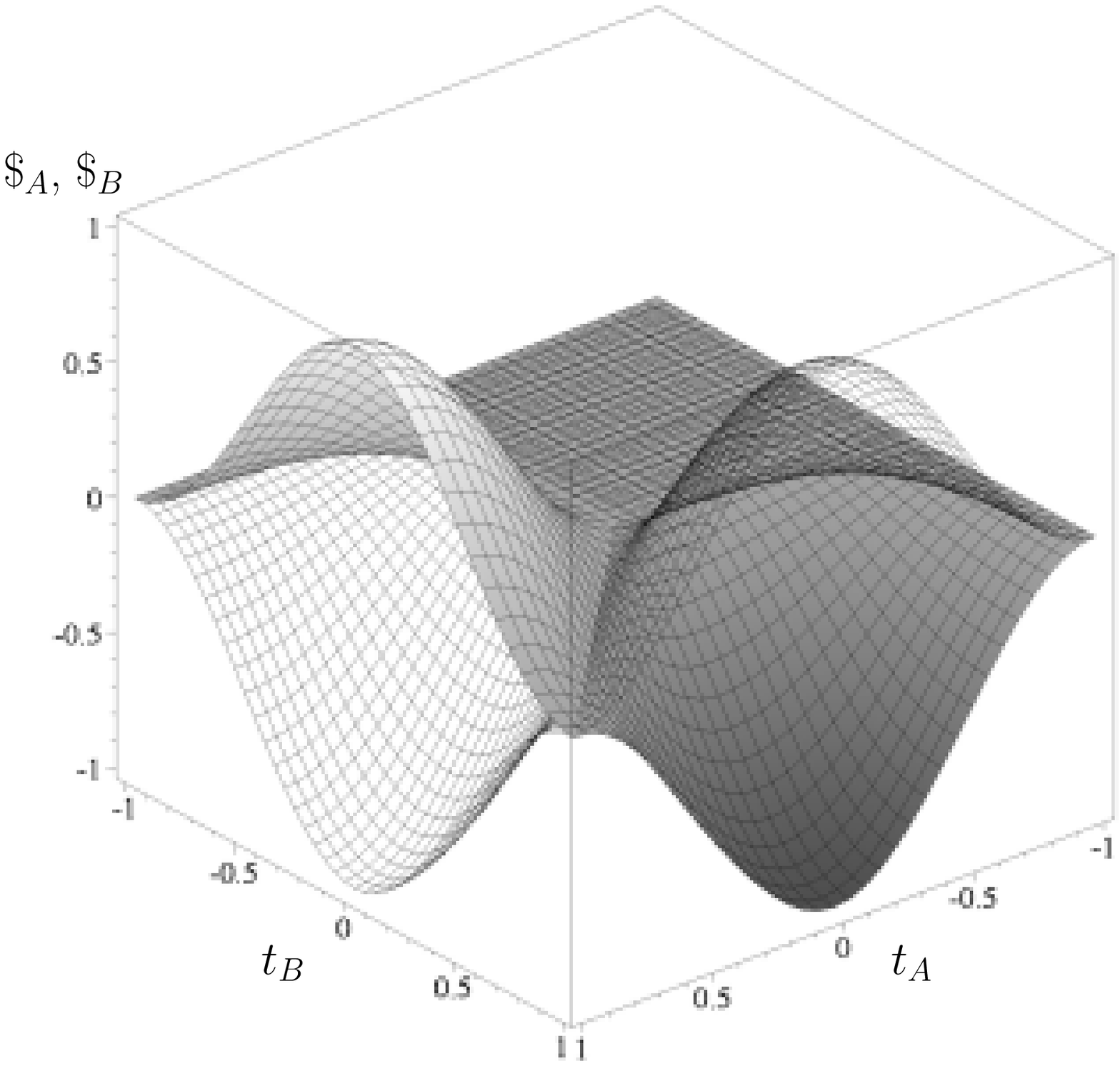}\vspace*{-4.7cm}\figsubcap{b}}
  \caption{Expected payoff of scientists A and B in a maximally entangled quantum game ((a): $\gamma=\frac{\pi}{2}$) and in a partially entangled quantum game ((b): $\gamma=\frac{\pi}{4}$). Payoff setting see Table \ref{tab:PayOff_case1}.}%
  \label{fig:1ent}
\end{center}
\end{figure*}
In Fig. \ref{fig:1ent}(a) the expected payoff for scientist A ($\$_A$, intransparent surface) and scientist B ($\$_B$, wired surface) is visualized; in contrast to Fig. \ref{fig:1sep} the players are maximally entangled ($\gamma=\frac{\pi}{2}$). Because of the change in the payoff surfaces, the strategy (\o,\o) is neither a dominant strategy nor a Nash equilibrium any more. For example, if player B chooses the strategy \o, it would be advisable for player A to select the strategy $\hat{{\cal{U}}}_A(0,\pi/2) \, \hat{=} \, (t_A=-1)$. In contrast to the disappearance of the former Nash equilibrium (\o,\o), new Nash equilibria are observed in the maximally entangled game. The pure quantum strategy $\hat{{\cal{Q}}}:=\hat{{\cal{U}}}(0,\pi/2) \, \hat{=} \, (t=-1)$ for instance is a Nash equilibrium because of the following conditions:
\begin{eqnarray}
&\$_A(t_A=-1,t_B=-1) =0 \geq& \nonumber\\
&-{\mbox{sin}\left(\frac{\theta_A}{2}\right)}^2 = \$_A(\hat{{\cal{U}}}_A,t_B=-1)
\quad \forall \,\,\theta_A \in [0,\pi] \quad ,& \nonumber\\ 
&\$_B(t_A=-1,t_B=-1) =0 \geq& \nonumber\\
&-{\mbox{sin}\left(\frac{\theta_B}{2}\right)}^2 = \$_B(t_B=-1,\hat{{\cal{U}}}_A)
\quad \forall \,\,\theta_B \in [0,\pi] \quad .& \nonumber
\end{eqnarray}
By examining Fig. \ref{fig:1ent}(a) one can see that all quantum strategies with $t\leq-0.5$ belong to the set of possible Nash equilibria. 

The results of the maximally entangled game show, that if quantum strategies are allowed, the scientists are not longer trapped in the strategy set (\o,\o). Nash equilibria exist only if both players choose a quantum strategy with $t_A,t_B\leq-0.5$. 

For partially entangled situations ($0<\gamma<\frac{\pi}{2}$), a boundary entanglement $\gamma_1=\frac{\pi}{4}$ can be specified, where the Nash equilibrium (\o,\o) fades to the quantum equilibria $t_A,t_B\leq-0.5$. Fig. \ref{fig:1ent}(b) depicts the partially entangled quantum game, which is right at the edge of dissolving the Nash equilibrium (\o,\o). 
For all $\gamma\leq\frac{\pi}{4}$ the Nash equilibrium of the game is (\o,\o), whereas for $\gamma>\frac{\pi}{4}$ the outcome of the game is similar to the maximally entangled situation, although the range of the set of quantum Nash equilibria is smaller and varies from ($\gamma=\frac{\pi}{4}$: $t_A,t_B = -1$) to ($\gamma=\frac{\pi}{2}$: $-1\leq (t_A,t_B) \leq-0.5$).  

\subsubsection{The Open Access Quantum Game as a Prisoners' Dilemma}
\begin{figure}[H]
\vspace*{0.25cm}
\centerline{
\includegraphics[width=3.5in]{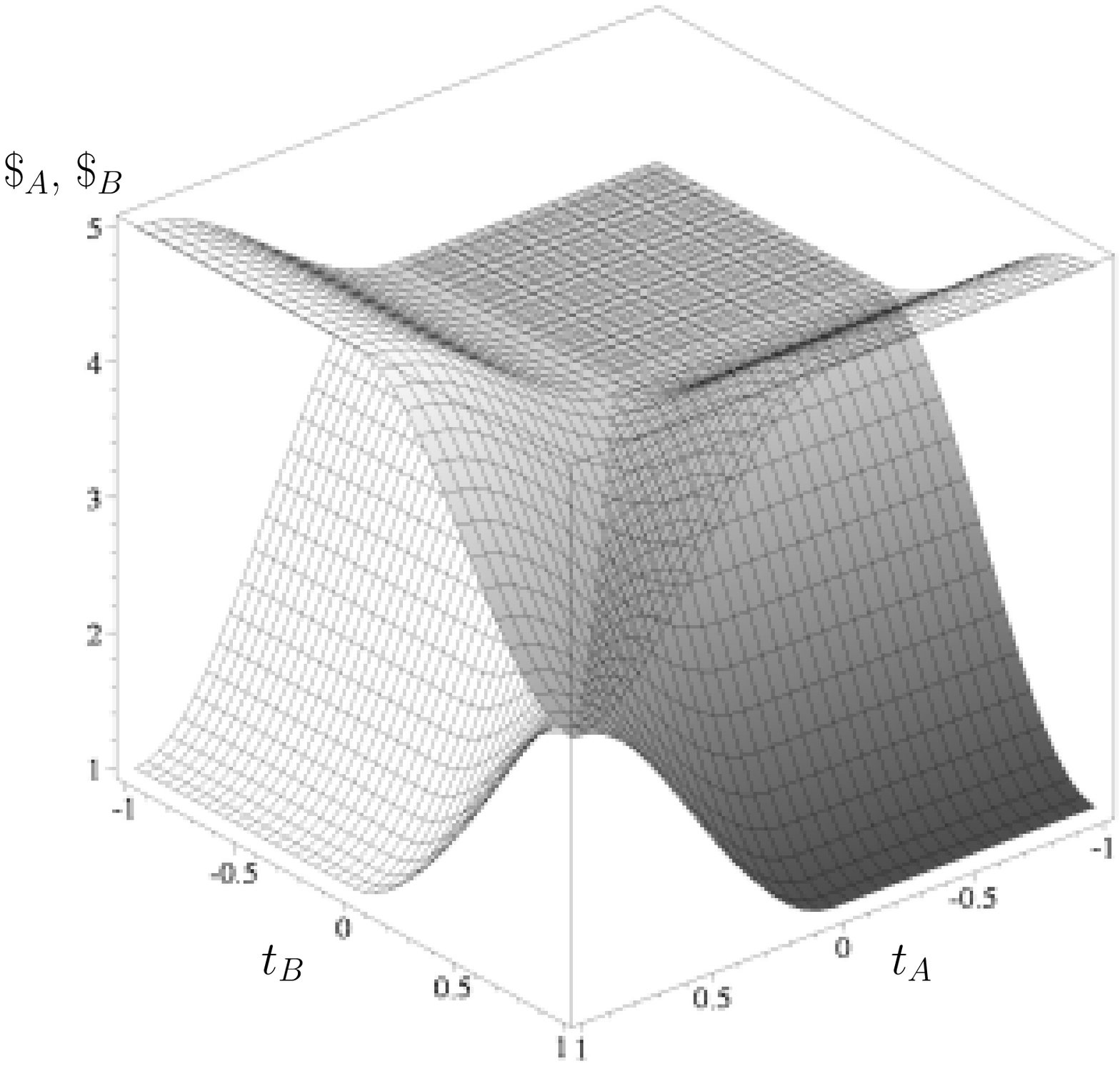}
}
\vspace*{-5.7cm}
\caption{Expected payoff of scientists A and B in a separable prisoners' dilemma quantum game (payoff setting see Table \ref{tab:PayOff_case2}).}
\label{fig:2sep}
\end{figure}

We now focus on an open access game with a payoff matrix similar to a prisoners' dilemma (see Table \ref{tab:PayOff_case2}). In difference to the zero sum game, discussed in the previous subsection, a dilemma occurs for both scientists. The players again are imprisoned in the strategy set (\o,\o), although a choice of (o,o) would be better for both of them. Fig. \ref{fig:2sep} illustrates this quandary in a graphic way (separable game with $\gamma=0$).
In contrast to Fig. \ref{fig:1sep}, where the strategy sets (o,o) and (\o,\o) are on the same payoff level ($\$_A(\mbox{o},\mbox{o})=\$_A(\mbox{\o},\mbox{\o})=0$; same for player B), the payoff magnitudes are now different ($\$_A(\mbox{o},\mbox{o})=4$, $\$_A(\mbox{\o},\mbox{\o})=3$; same for player B). The plane of the quantum-quantum region in Fig. \ref{fig:2sep} ($t_A,t_B\leq0$) has moved upwards and has a higher payoff than the dominant strategy (\o,\o). There is again no difference between the classical outcome of the game and the separable quantum version: (\o,\o) remains to be a dominant strategy. 

\begin{figure}
\vspace*{0.25cm}
\centerline{
\includegraphics[width=3.5in]{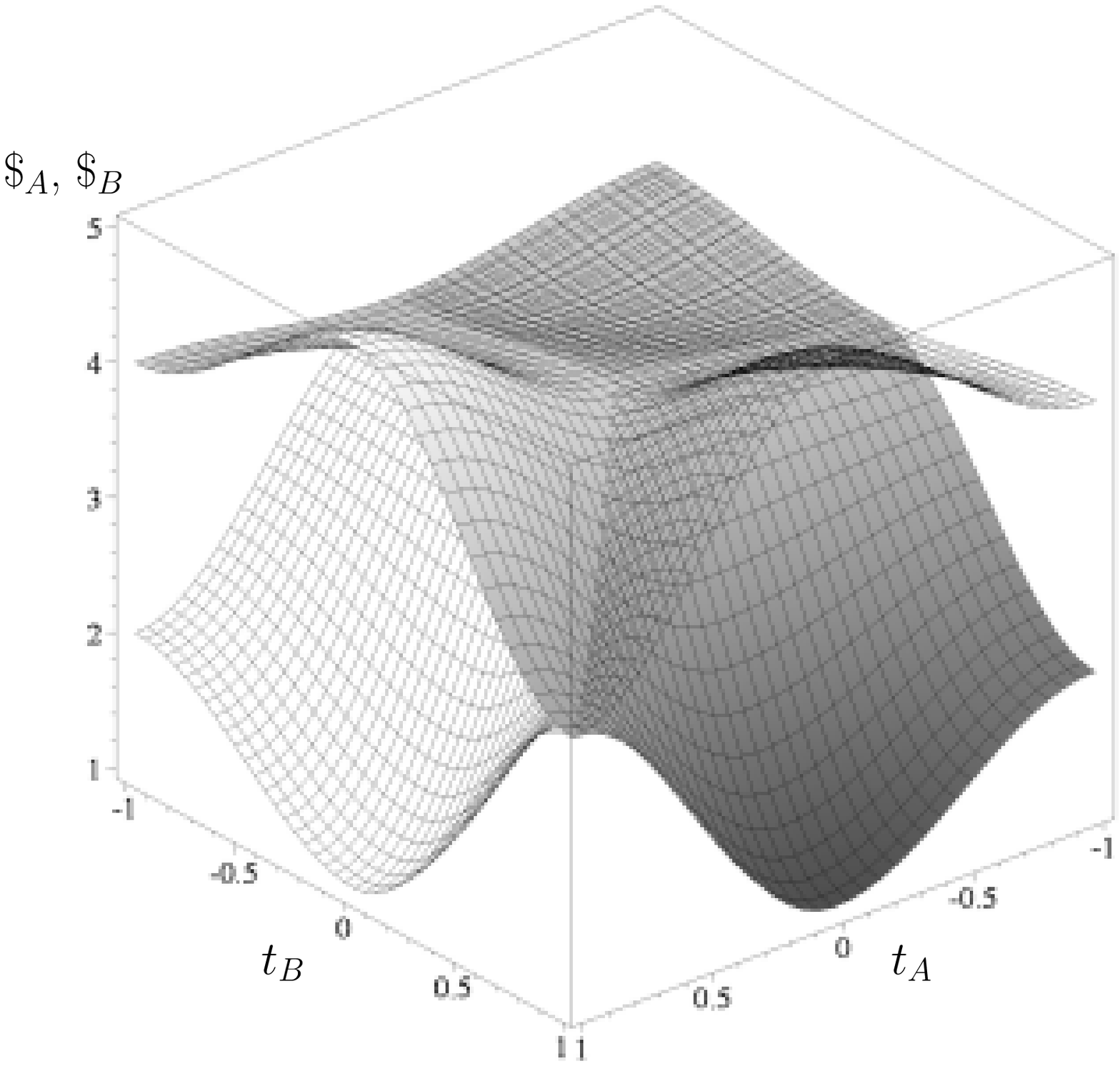}
}
\vspace*{-5.7cm}
\caption{Expected payoff of scientists A and B in partially entangled prisoners' dilemma quantum game (payoff setting see Table \ref{tab:PayOff_case2}, $\gamma=2\, \mbox{arctan}(\frac{\sqrt{3}-1}{\sqrt{3}+1})$).}
\label{fig:2entab_a}
\end{figure}

\begin{figure}
\vspace*{0.25cm}
\centerline{
\includegraphics[width=3.5in]{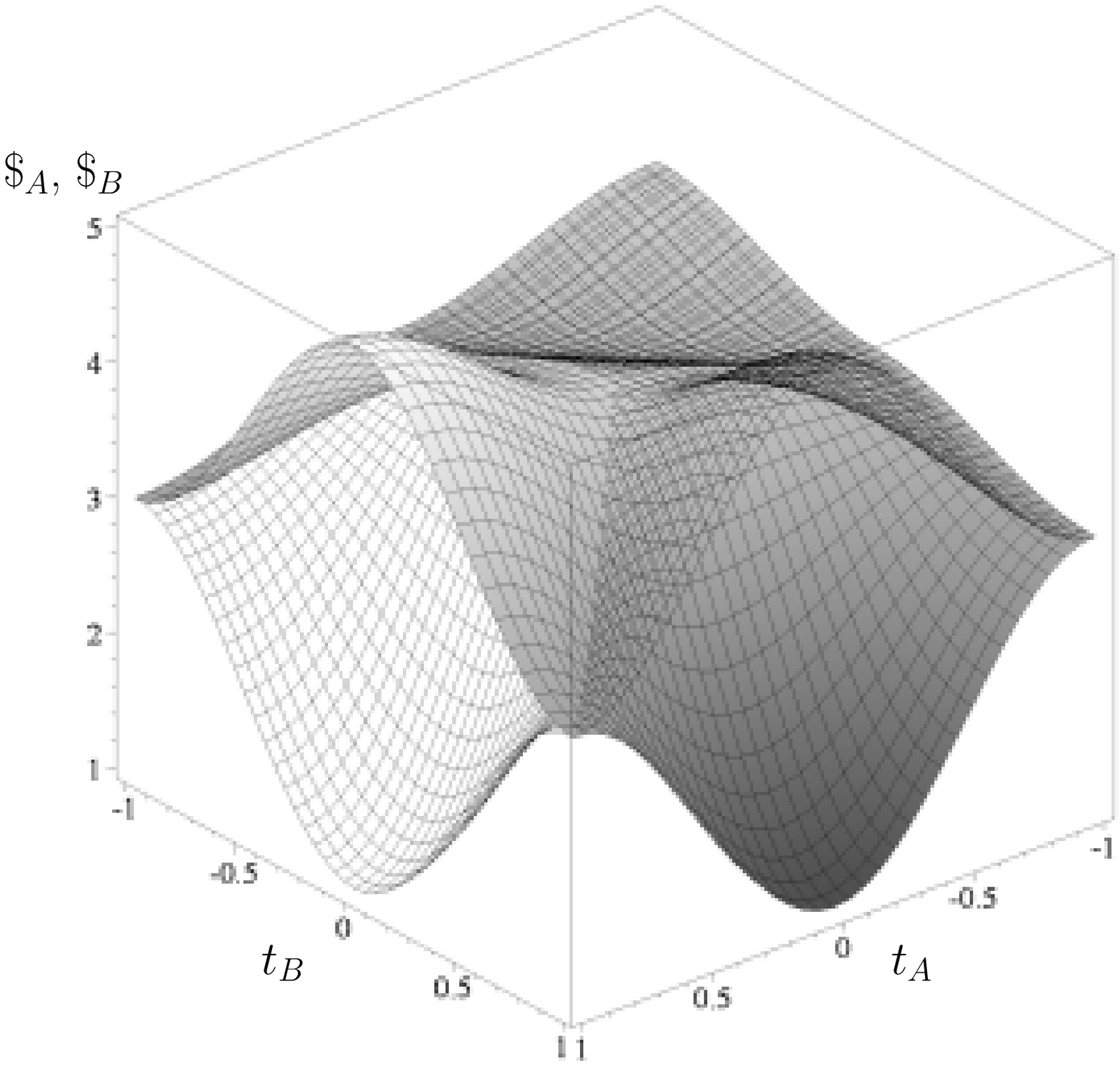}
}
\vspace*{-5.7cm}
\caption{Expected payoff of scientists A and B in partially entangled prisoners' dilemma quantum game (payoff setting see Table \ref{tab:PayOff_case2}, $\gamma=\frac{\pi}{4}$).}
\label{fig:2entab_b}
\end{figure}

\begin{figure}
\vspace*{0.25cm}
\centerline{
\includegraphics[width=3.5in]{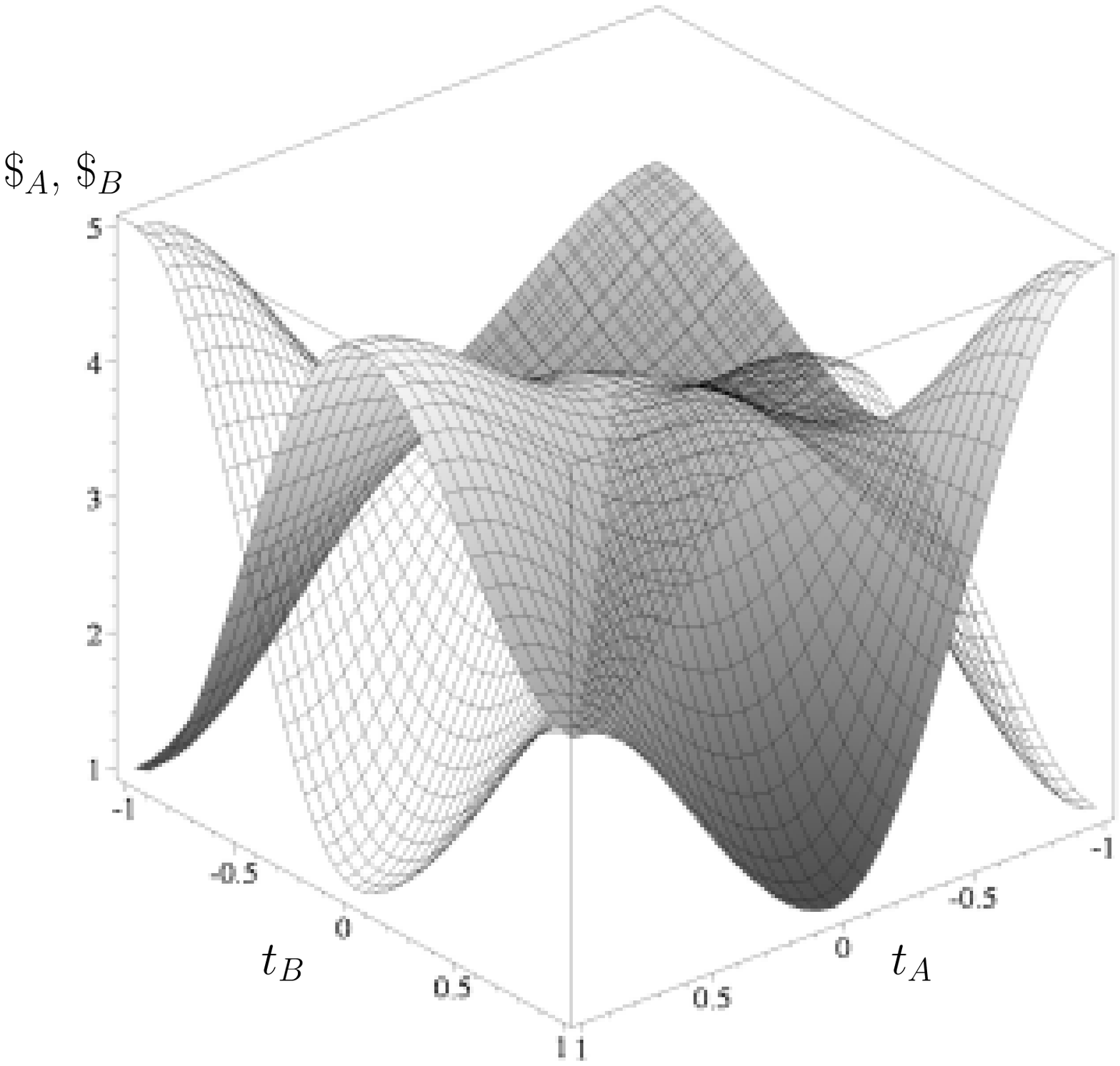}
}
\vspace*{-5.7cm}
\caption{Expected payoff of scientists A and B in a maximally entangled prisoners' dilemma quantum game (payoff setting see Table \ref{tab:PayOff_case2}).}
\label{fig:2ent}
\end{figure}

Increasing the entanglement factor $\gamma$ to higher values leads to a qualitative change in the outcome of the game, if its value overruns $\gamma_1:=2\, \mbox{arctan}(\frac{\sqrt{3}-1}{\sqrt{3}+1})$. For $\gamma_1<\gamma$ the strategy (\o,\o) ceases to be a unique dominant strategy, however (\o,\o) remains to be a Nash equilibrium if the entanglement-factor lies in the range $\gamma_1<\gamma\leq\gamma_2:=\frac{\pi}{4}$. In this range, there exist two Nash equilibria, namely the former Nash equilibrium (\o,\o) and a new quantum Nash equilibrium ($\hat{{\cal{Q}}}_A,\hat{{\cal{Q}}}_B$), which corresponds to ($t_A=-1$, $t_B=-1$). Fig. \ref{fig:2entab_a} shows the payoff surfaces for both players at the entanglement barrier $\gamma_1$. 
If one further increases $\gamma$, the strategy (\o,\o) even ceases to be a Nash equilibrium. For example, if $\gamma>\gamma_2$ and player B chooses the strategy \o, the best reward for player A would be the quantum strategy $\hat{{\cal{Q}}}_A$. Fig. \ref{fig:2entab_b} depicts the payoff surfaces for both players for $\gamma=\gamma_2$. 
For $\gamma>\gamma_2$ there exists only the quantum Nash equilibrium ($\hat{{\cal{Q}}}_A,\hat{{\cal{Q}}}_B$), as one can see by looking at the maximally entangled situation (Fig. \ref{fig:2ent}).

It should be mentioned, that our results are different from the results presented in \cite{eisert-1999-83} and \cite{du-2002-88}, which is due to a different payoff matrix. For the separable and maximally entangled game there is no qualitative difference in the outcomes, whereas we want to point out, that we find different Nash equilibria for the partially entangled games (see Fig. \ref{fig:2entab_a}, \ref{fig:2entab_b}). J. Du et al. found the two Nash equilibria (($\hat{{\cal{Q}}}$,\o) and (\o,$\hat{{\cal{Q}}}$)) for $\mbox{arcsin}(\sqrt{\frac{1}{5}})<\gamma\leq\mbox{arcsin}(\sqrt{\frac{2}{5}})$ \cite{du-2002-88}, which is in clear contrast to our results. We therefore want to emphasize, that if one extends a prisoners' dilemma into a quantum region, the structure of the payoff matrix is important and seems to separate different types of quantum prisoners' dilemmas when varying the systems' entanglement.

\begin{figure*}%
\vspace*{1.35cm}
\begin{center}
  \parbox{3.1in}{\includegraphics[width=3.0in]{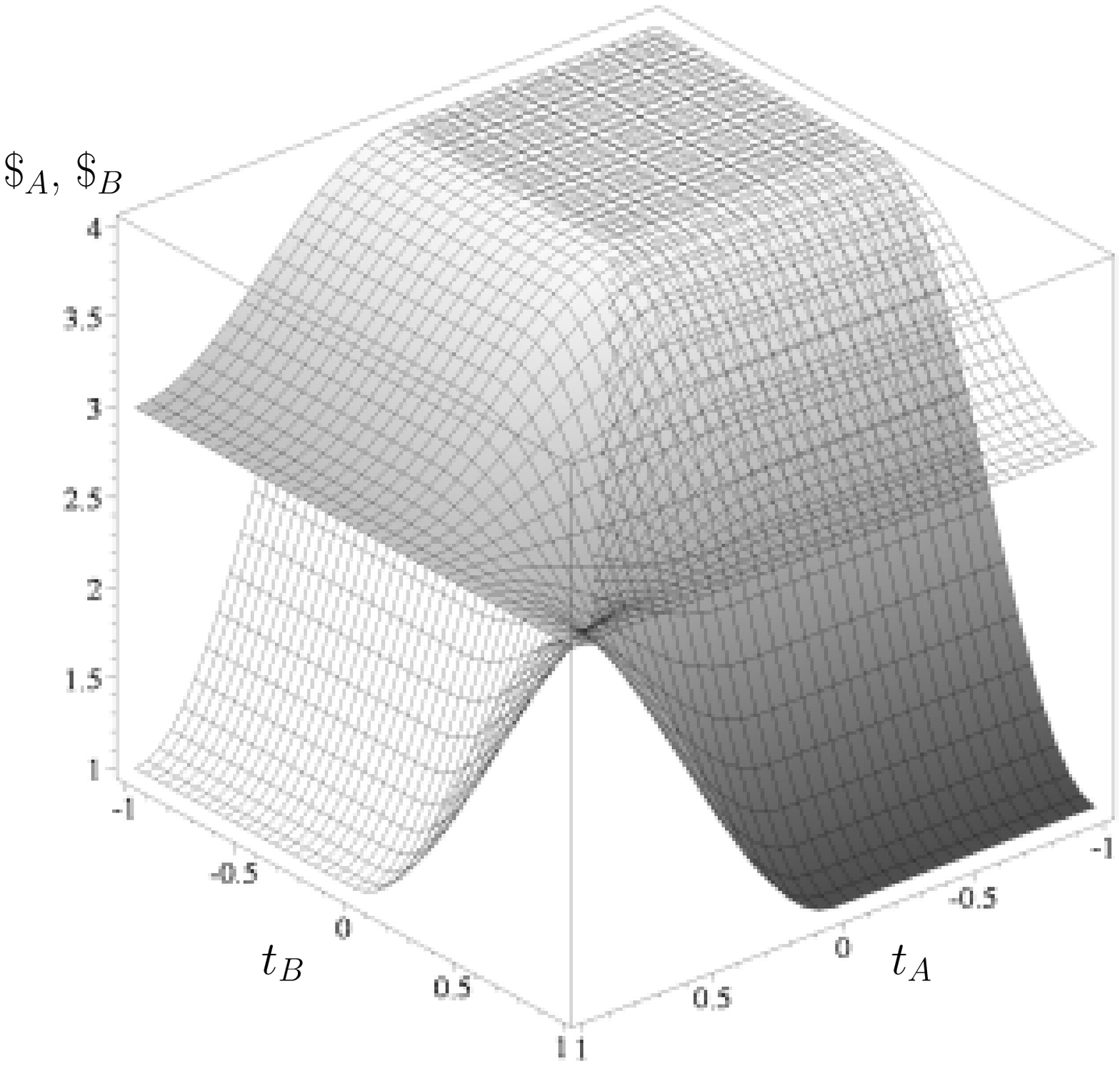}\vspace*{-4.7cm}\figsubcap{a}}
  \parbox{3.1in}{\includegraphics[width=3.0in]{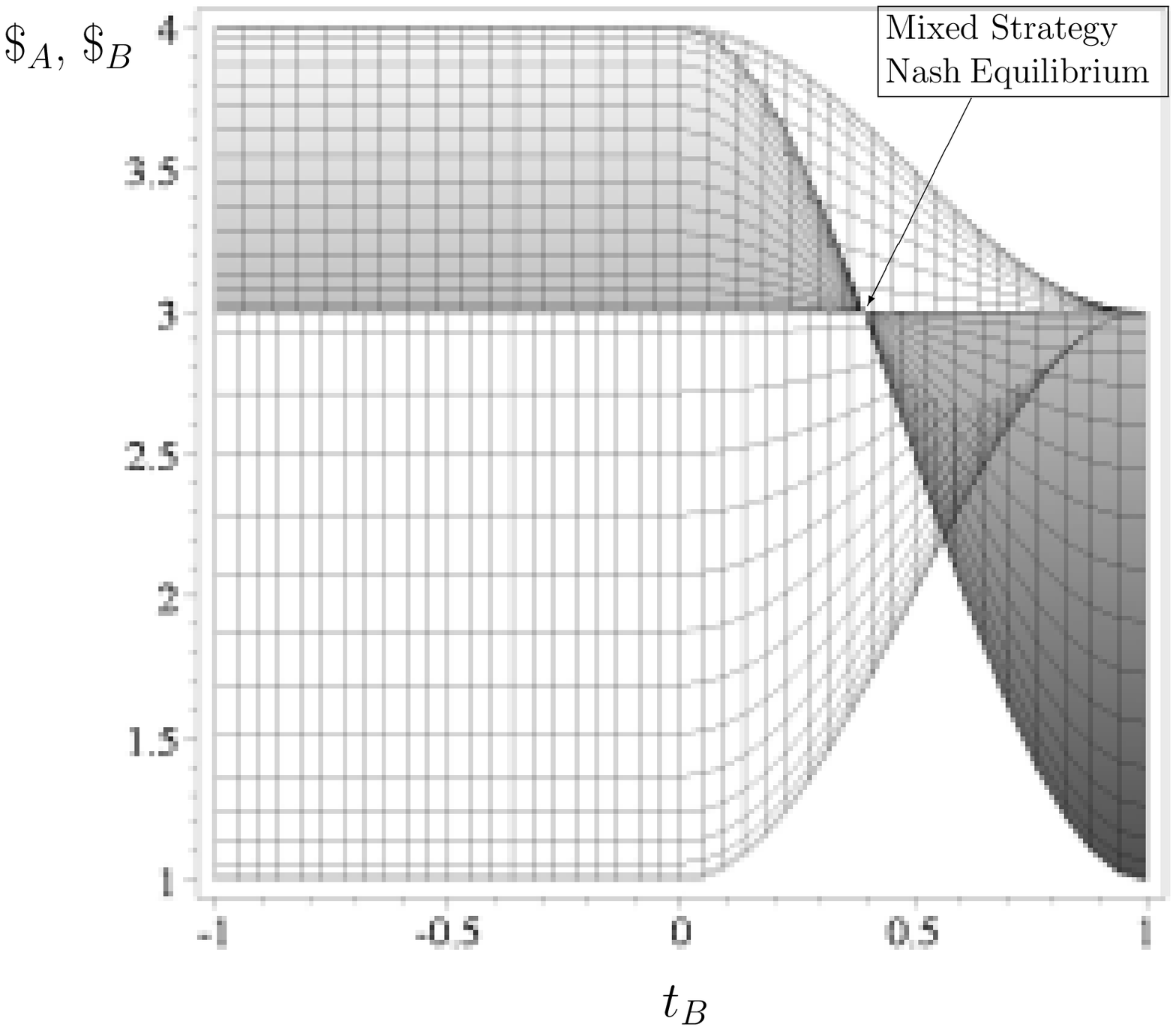}\vspace*{-4.7cm}\figsubcap{b}}
  \caption{(a) shows the expected payoff of scientists A and B in a separable stag hunt quantum game (payoff setting see Table \ref{tab:PayOff_case3}). (b) depicts the projection of figure (a) onto the $\$$-$t_B$ plane.}%
  \label{fig:3sep}
\end{center}
\end{figure*}
\subsubsection{Open Access as a Stag Hunt Quantum Game}
In contrast to the other separable games discussed in the previous subsections, the stag hunt quantum version of the open access game even shows advantages of using quantum strategies in the separable situation, where the strategical operations of the scientists are not entangled. In this case the QQ-plane of the payoffs for both players always lies above or equal to all other payoff values (see Fig. \ref{fig:3sep}(a)). In addition to the three classical Nash equilibria ((\o,\o), (o,o) and $\frac{2}{3}$(o,o)), a set of new quantum Nash equilibria can be observed within the separable quantum game ($t_A,t_B<0$). All quantum strategies that lie on the QQ-plane of Fig. \ref{fig:3sep}(a), ensure an identical, rather high payoff for both players ($\$_A(QQ)=\$_B(QQ)=4$). Because of the absence of a dominant strategy and the complex structure of Nash equilibria, it is difficult to predict the outcome of the game. A risk conducted player may prefer the strategy \o, because this will guarantee him a payoff of 3. A payoff conducted player might be guided by the possibility of getting a greater payoff, and therefore will prefer either strategy o, or a quantum strategy $t<0$. The mixed strategy Nash equilibrium $\frac{2}{3}$(o,o) can be visualized if one examines the surfaces from a viewpoint parallel to the strategy space of player A (see Fig. \ref{fig:3sep}(b)). The character of a mixed Nash equilibrium ($t_A^\star$, $t_B^\star$) is that the gradients of the payoff surfaces vanish:
\begin{eqnarray}
\left.\frac{\partial \, \$_A}{\partial \, t_A}(t_A,t_B)\right|_{t_B=t_B^\star}&\equiv& 0 \, , \quad \forall \,\, t_A \in [-1,1]\\
\left.\frac{\partial \, \$_B}{\partial \, t_B}(t_A,t_B)\right|_{t_A=t_A^\star}&\equiv& 0 \, , \quad \forall \,\,  t_B \in [-1,1] \quad .\nonumber
\end{eqnarray}
$t_B^\star$ for example can be observed in the special projection of Fig. \ref{fig:3sep}(b), where the whole payoff surface of player A ($\$_A$) contracts to one single point. From our calculations we get the following mixed strategy Nash equilibrium ($t^\star=t_A^\star=t_B^\star=\frac{2}{\pi}\mbox{arcsin}(\frac{1}{\sqrt{3}})$), which corresponds to the strategy $\frac{2}{3}$(o,o).

\begin{figure}[H]
\vspace*{0.25cm}
\centerline{
\includegraphics[width=3.5in]{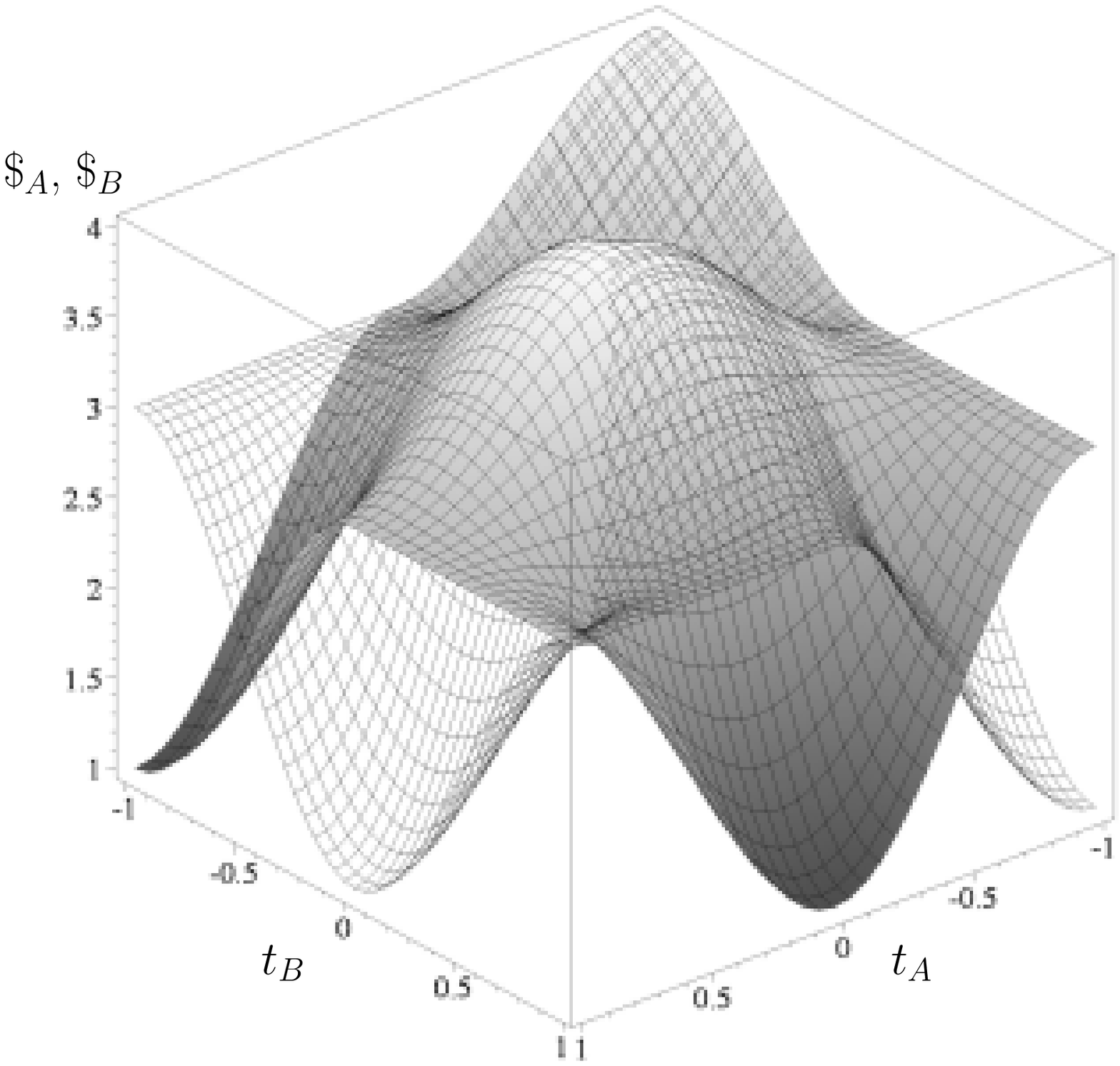}
}
\vspace*{-5.7cm}
\caption{Expected payoff of scientists A and B in a maximally entangled stag hunt quantum game (payoff setting see Table \ref{tab:PayOff_case3}).}
\label{fig:3ent}
\end{figure}

The maximally entangled stag hunt quantum game is displayed in Fig. \ref{fig:3ent}.  
\begin{figure*}
\vspace*{0.4cm}
\begin{center}
  \parbox{3.3in}{\includegraphics[width=3.2in]{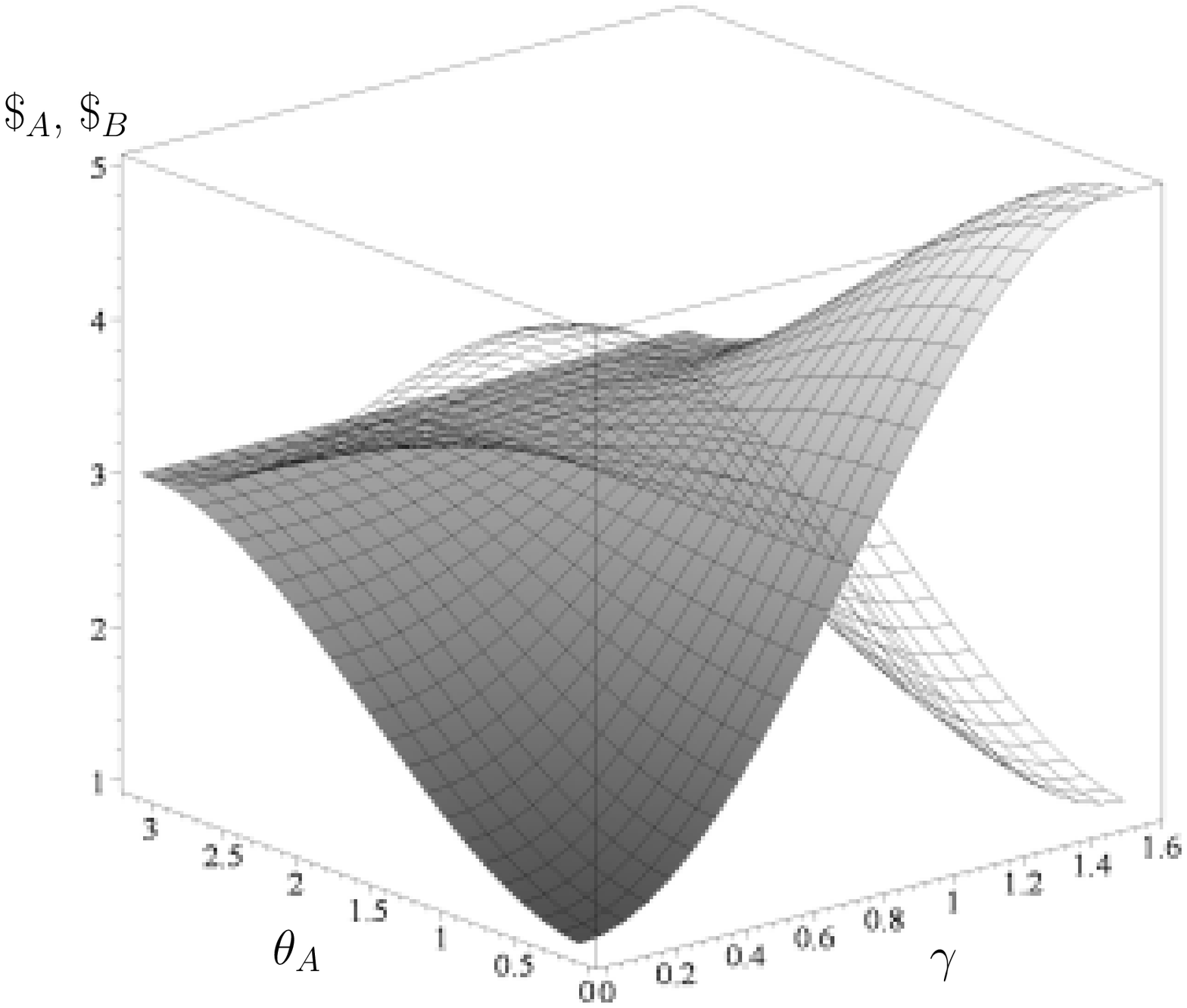}\vspace*{-4.7cm}\figsubcap{a}}
  \parbox{3.3in}{\includegraphics[width=3.2in]{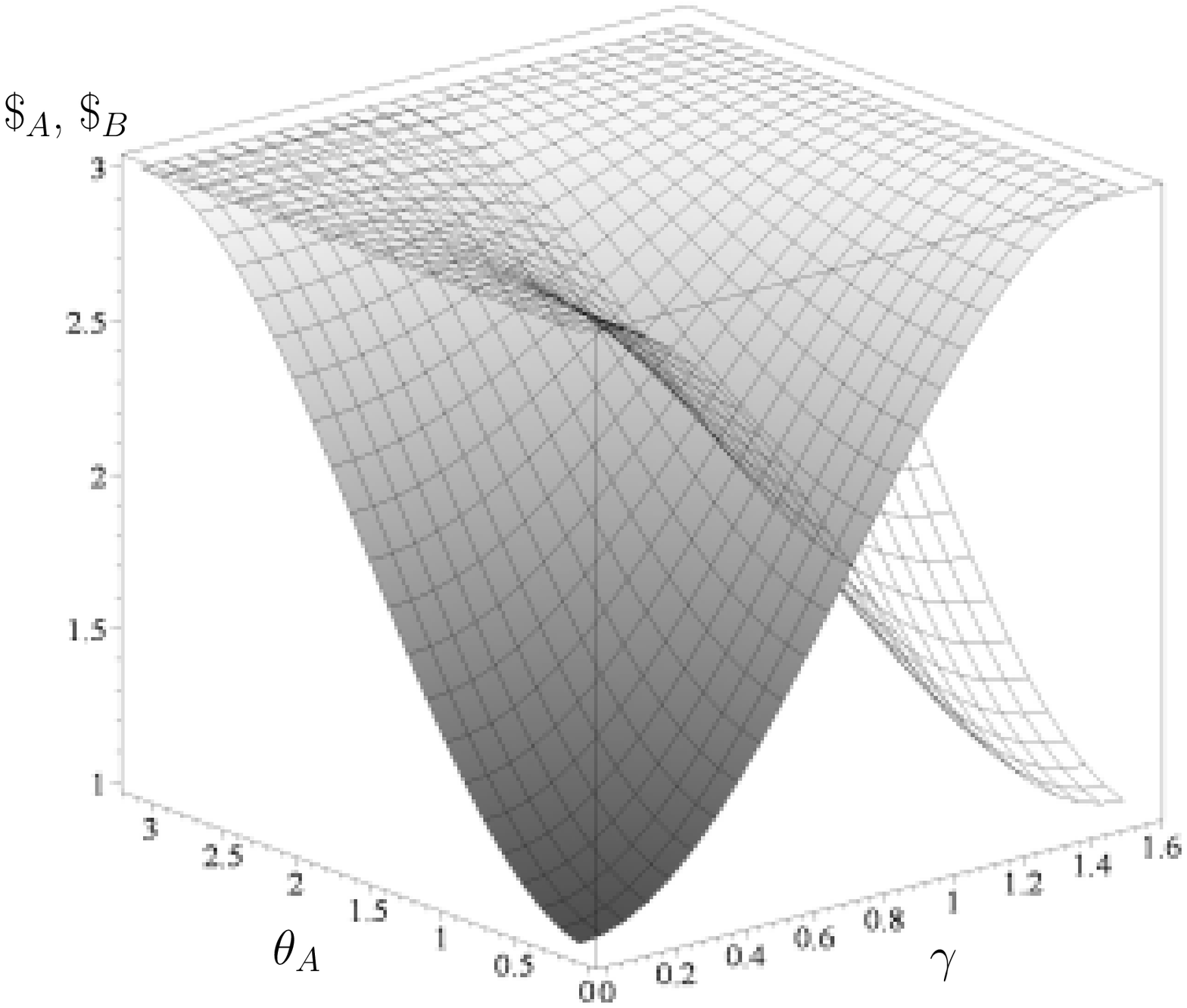}\vspace*{-4.7cm}\figsubcap{b}}
  \caption{Expected payoff of scientists A and B versus $\theta_A$ and $\gamma$. Player B has selected the classical strategy {\o}, whereas player A selects a quantum strategy $\hat{{\cal{U}}}_A = \hat{{\cal{U}}}(\theta_A,\frac{\pi}{2})$. (a) shows the prisoners' dilemma case whereas (b) depicts the stag hunt quantum game.}%
  \label{fig:gamma}
\end{center}
\end{figure*}

In this version of the game three Nash equilibria occur, namely (\o,\o), $\frac{2}{3}$(o,o) and ($\hat{{\cal{Q}}}_A,\hat{{\cal{Q}}}_B$). Although (\o,\o) technically  remains to be a Nash equilibrium, no rational acting player would choose such a strategy, because the alternative of the quantum strategy $\hat{{\cal{Q}}}$ would give him in any case a better or equal payoff: 
\begin{eqnarray}
&\mbox{Quantum Strategy:}&\nonumber\\
&\$_A(\hat{{\cal{Q}}}_A,t_B) \geq 3 \wedge \$_B(t_A,\hat{{\cal{Q}}}_B) \geq 3 
\quad \forall \,\, t_A,t_B \in [-1,1]& \nonumber\\
&\mbox{Non-Open Access:}&\nonumber\\
&\$_A(\mbox{\o},t_B) \leq 3 \wedge \$_B(t_A,\mbox{\o}) \leq 3 
\quad \forall \,\, t_A,t_B \in [-1,1]& \nonumber
\end{eqnarray}
Furthermore it should be mentioned, that for all types of entanglement the mixed strategy Nash equilibrium $\frac{2}{3}$(o,o) persists at its former position. 

In summary, we conclude that the players of a maximally entangled stag hunt quantum game will be in favor of performing the quantum strategy $\hat{{\cal{Q}}}$ over the non-open access strategy \o. 

\subsection{Manifestation of Quantum Strategies}
We want to point out, that the measurable choice of the quantum strategy $\hat{{\cal{Q}}}$ in reality does not necessarily appear as the strategy o -- albeit, if both players will choose $\hat{{\cal{Q}}}$, the measured outcome will be (o,o). To illustrate the role of entanglement and the nature of quantum strategies, we have fixed the strategy of scientist B to $\hat{{\cal{U}}}_B = \hat{{\cal{U}}}(\pi,0)=\mbox{\o}$, whereas we choose the strategy of scientist A to be a quantum strategy $\hat{{\cal{U}}}_A = \hat{{\cal{U}}}(\theta_A,\frac{\pi}{2})$. Fig. \ref{fig:gamma} displays the payoff for the players A and B as a function of $\theta_A$ and $\gamma$. Fig. \ref{fig:gamma}(a) depicts the calculations for the prisoners' dilemma game, whereas Fig. \ref{fig:gamma}(b) shows the results within the stag hunt quantum game. If the scientists' strategies are not entangled ($\gamma = 0$), the best respond for player A in the prisoners' dilemma game is the choice of $\theta_A=\pi$, which would result in the classical Nash equilibrium (\o,\o), giving both players the payoff 3. In contrast, if we focus on a situation where the scientists' strategies are maximally entangled ($\gamma = \frac{\pi}{2}$), the best respond for scientist A is $\theta_A=0$, giving him a payoff of 5 and player B a payoff of 1. Player B could be amazed about his little payoff. To understand the real cause, we need to examine the joint probabilities of the measurable outcomes of the game. If player B selects the classical strategy {\o} and player A chooses the quantum strategy $\hat{{\cal{Q}}}$, the joint probabilities result in the following outcomes:
\begin{eqnarray}
&\left| \, \left< \mbox{{o}{o}} | \Psi_f \right> \, \right|^2 \, = \, 
\left| \, \left< \mbox{{\o}{\o}} | \Psi_f \right> \, \right|^2 \, = \, 0 \quad ,&  \\
&\left| \, \left< \mbox{{o}{\o}} | \Psi_f \right> \, \right|^2 \, = \, 
{\mbox{cos}\left( \gamma \right)}^2 \, , \quad
\left| \, \left< \mbox{{\o}{o}} | \Psi_f \right> \, \right|^2 \, = \, 
{\mbox{sin}\left( \gamma \right)}^2 &\quad .\nonumber
\end{eqnarray}

\begin{figure}
\vspace*{0.0cm}
\centerline{
\includegraphics[width=4.0in]{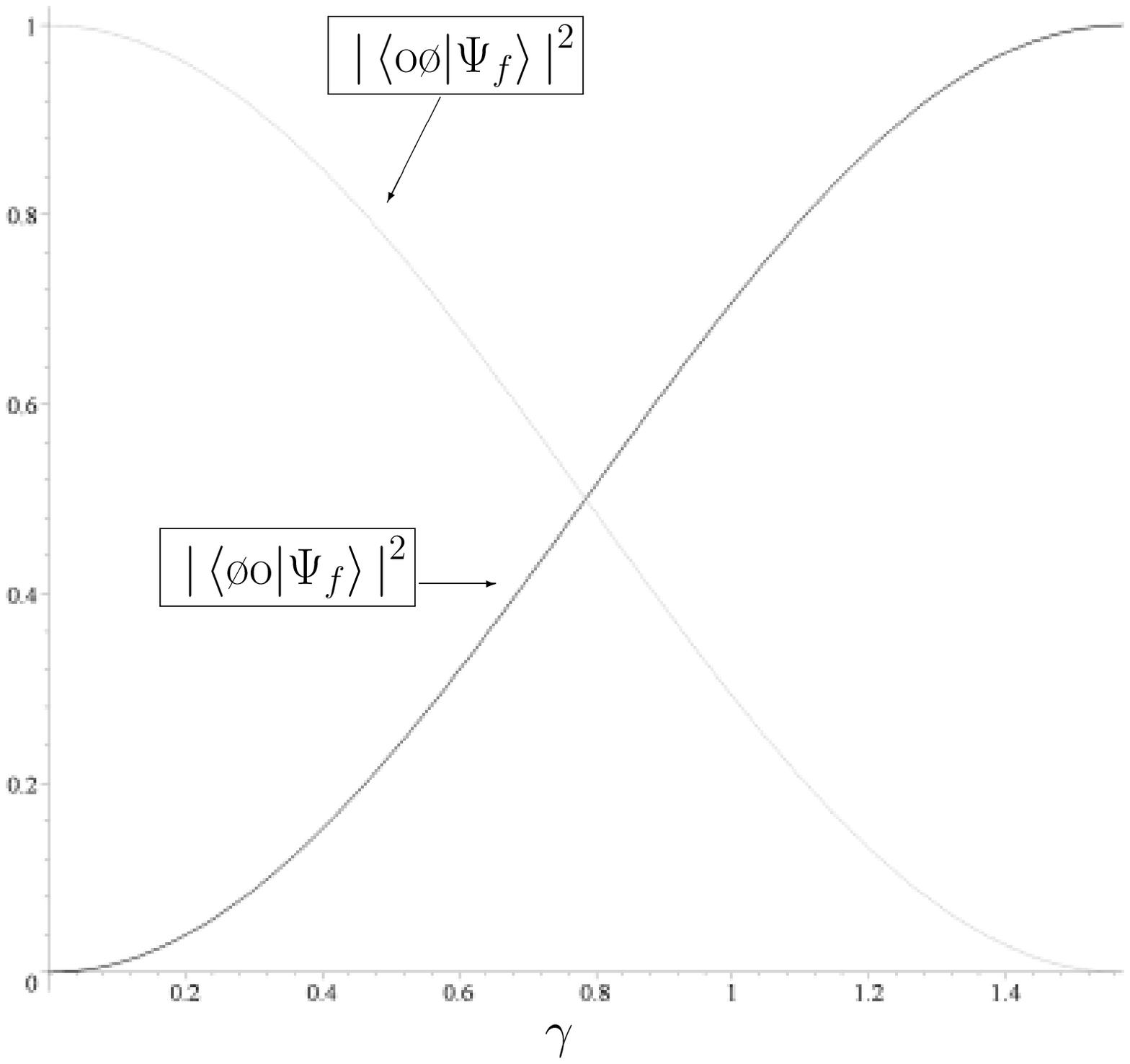}
}
\vspace*{-6.0cm}
\caption{Joint probabilities of the measurable outcomes as a function of $\gamma$. Player B chooses strategy {\o}, whereas Player A chooses $\hat{{\cal{Q}}}$.}
\label{fig:sum1}
\end{figure}

In Fig. \ref{fig:sum1} the non-zero probabilities $\left| \, \left< \mbox{{o}{\o}} | \Psi_f \right> \, \right|^2$ and $\left| \, \left< \mbox{{\o}{o}} | \Psi_f \right> \, \right|^2$ are plotted against the scientists' entanglement $\gamma$. 
The cause of the amazement of player B is that even though he chooses the strategy {\o}, the probability of measuring {\o} is zero if the entanglement $\gamma$ is maximal. By using the quantum strategy $\hat{{\cal{Q}}}$ player A is able to switch the choice of player B. Within an entangled quantum game, it is not feasible to insist on a classically chosen strategy. 

\section{Summary}\label{sec4}
This article focuses the question why the open access model is only successfully adopted by a few scientific disciplines. We have constructed a game theoretical model, where the scientists' incentives where described with a reputation dependent payoff matrix. Three game settings where addressed, namely a zero sum game, the prisoners' dilemma and a stag hunt version of the open access game. By calculating the outcome of the games within a classical game theoretical framework, we have shown that in all cases the scientists face a dilemma situation: Considering a potential loss in reputation, incentives to perform open access are missing. These findings change, if quantum strategies are allowed. If the entanglement overruns a certain barrier, quantum strategies become superior to the former Nash equilibrium strategies. In none of the three different game settings the choice of traditional publishing remains to be a rational strategy for the players, if their strategical choices are maximally entangled. The results of this article therefore indicate one possible explanation of the differing publishing methods of scientific communities. In quantum game theory parlance one would say, that scientific disciplines, like mathematics and physics, which had been successful in realizing the open access model, consist of scientists, whose strategical operations are strongly entangled. In contrast, if a scientific community is still imprisoned in the Nash equilibrium of non-open access, there would be a lack of entanglement between the strategical choices of the related scientists of the community.\\ 

\section*{Acknowledgments}
We want to thank Jens Eisert for helpful discussions.
This research is supported by grants from the German National Science Foundation (DFG) (Project “Scientific Publishing and Alternative Pricing Mechanisms”, Grant No. GZ 554922). We gratefully acknowledge the financial support.


\begin{thebibliography}{28}
\expandafter\ifx\csname natexlab\endcsname\relax\def\natexlab#1{#1}\fi
\expandafter\ifx\csname bibnamefont\endcsname\relax
  \def\bibnamefont#1{#1}\fi
\expandafter\ifx\csname bibfnamefont\endcsname\relax
  \def\bibfnamefont#1{#1}\fi
\expandafter\ifx\csname citenamefont\endcsname\relax
  \def\citenamefont#1{#1}\fi
\expandafter\ifx\csname url\endcsname\relax
  \def\url#1{\texttt{#1}}\fi
\expandafter\ifx\csname urlprefix\endcsname\relax\def\urlprefix{URL }\fi
\providecommand{\bibinfo}[2]{#2}
\providecommand{\eprint}[2][]{\url{#2}}

\bibitem[{\citenamefont{Getz}(1999)}]{getz-1999}
\bibinfo{author}{\bibfnamefont{M.}~\bibnamefont{Getz}},
  \bibinfo{journal}{Proceedings of the ACRL 9th National Conference 1999}
  (\bibinfo{year}{1999}).

\bibitem[{\citenamefont{McCabe}(2004)}]{mccabe-2004}
\bibinfo{author}{\bibfnamefont{M.}~\bibnamefont{McCabe}},
  \bibinfo{journal}{Contributions to Economic Analysis and Policy}
  \textbf{\bibinfo{volume}{3}}, \bibinfo{pages}{1} (\bibinfo{year}{2004}).

\bibitem[{\citenamefont{SQW}(2003)}]{wellcome_2003}
\bibinfo{author}{\bibnamefont{SQW}}, \emph{\bibinfo{title}{Economic Analysis of
  Scientific Research Publishing.}} (\bibinfo{publisher}{Wellcome Trust,
  Cambridgeshire}, \bibinfo{year}{2003}).

\bibitem[{\citenamefont{Tenopir and King}(2000)}]{tenopir_2000}
\bibinfo{author}{\bibfnamefont{C.}~\bibnamefont{Tenopir}} \bibnamefont{and}
  \bibinfo{author}{\bibfnamefont{D.~W.} \bibnamefont{King}},
  \emph{\bibinfo{title}{Towards Electronic Journals.}} (\bibinfo{publisher}{SLA
  Publishing, Washington DC}, \bibinfo{year}{2000}).

\bibitem[{\citenamefont{Okerson}(1996)}]{okerson-1996}
\bibinfo{author}{\bibfnamefont{A.}~\bibnamefont{Okerson}},
  \bibinfo{journal}{Proceedings of the UNESCO Invitational Meeting on the
  Future of Scientific Information}  (\bibinfo{year}{1996}).

\bibitem[{\citenamefont{Tananbaum}(2003)}]{tananbaum-2003}
\bibinfo{author}{\bibfnamefont{G.}~\bibnamefont{Tananbaum}},
  \bibinfo{journal}{Learned Publishing} \textbf{\bibinfo{volume}{16(4)}},
  \bibinfo{pages}{284} (\bibinfo{year}{2003}).

\bibitem[{dug(2004)}]{dugall-2004}
\emph{\bibinfo{title}{5th Frankfurt Scientific Symposium: Is there any progress
  in alternative publishing?}} (\bibinfo{year}{2004}),
  \bibinfo{note}{\href{http://wiap.wiwi.uni-frankfurt.de/5thsymp/}{http://wiap.wiwi.uni-frankfurt.de/5thsymp/}}.

\bibitem[{\citenamefont{Harnad}(2005)}]{harnad-2005}
\bibinfo{author}{\bibfnamefont{S.}~\bibnamefont{Harnad}},
  \bibinfo{journal}{Ariadne} \textbf{\bibinfo{volume}{42}}
  (\bibinfo{year}{2005}),
  \bibinfo{note}{\href{http://www.ariadne.ac.uk/issue42/harnad/intro.html}{http://www.ariadne.ac.uk/issue42/harnad/intro.html}}.

\bibitem[{\citenamefont{Guedon}(2004)}]{guedon-2004}
\bibinfo{author}{\bibfnamefont{J.}~\bibnamefont{Guedon}},
  \bibinfo{journal}{Serials Review} \textbf{\bibinfo{volume}{30 (4)}},
  \bibinfo{pages}{315} (\bibinfo{year}{2004}).

\bibitem[{\citenamefont{EU}(2006)}]{eu_2006}
\bibinfo{author}{\bibnamefont{EU}}, \emph{\bibinfo{title}{Study on the economic
  and technical evolution of the scientific publication markets in Europe.}}
  (\bibinfo{publisher}{DG Research}, \bibinfo{year}{2006}).

\bibitem[{\citenamefont{DFG}(2005)}]{dfg_2003}
\bibinfo{author}{\bibnamefont{DFG}}, \emph{\bibinfo{title}{Publication
  Strategies in Transformation?}} (\bibinfo{publisher}{Wiley-VCH Verlag
  (Weinheim)}, \bibinfo{year}{2005}),
  \bibinfo{note}{http://www.dfg.de/en/dfg\_profile/facts\_and\_figures/
  \\statistical\_reporting/open\_access/index.html}.

\bibitem[{\citenamefont{Schroter et~al.}(2005)\citenamefont{Schroter, Tite, and
  Smith}}]{schroter-2005}
\bibinfo{author}{\bibfnamefont{S.}~\bibnamefont{Schroter}},
  \bibinfo{author}{\bibfnamefont{L.}~\bibnamefont{Tite}}, \bibnamefont{and}
  \bibinfo{author}{\bibfnamefont{R.}~\bibnamefont{Smith}},
  \bibinfo{journal}{BMJ}  (\bibinfo{year}{2005}),
  \bibinfo{note}{doi:10.1136/bmj.38359.695220.82}.

\bibitem[{\citenamefont{Dewett and Denisi}(2004)}]{dewett-2004}
\bibinfo{author}{\bibfnamefont{T.}~\bibnamefont{Dewett}} \bibnamefont{and}
  \bibinfo{author}{\bibfnamefont{A.~S.} \bibnamefont{Denisi}},
  \bibinfo{journal}{Scientometrics} \textbf{\bibinfo{volume}{60}},
  \bibinfo{pages}{249} (\bibinfo{year}{2004}).

\bibitem[{\citenamefont{Lawrence}(2001)}]{lawrence-2001}
\bibinfo{author}{\bibfnamefont{S.}~\bibnamefont{Lawrence}},
  \bibinfo{journal}{Nature} \textbf{\bibinfo{volume}{411}},
  \bibinfo{pages}{521} (\bibinfo{year}{2001}), \bibinfo{note}{31 May}.

\bibitem[{\citenamefont{Harnad and Brody}(2004)}]{harnad-2004}
\bibinfo{author}{\bibfnamefont{S.}~\bibnamefont{Harnad}} \bibnamefont{and}
  \bibinfo{author}{\bibfnamefont{T.}~\bibnamefont{Brody}},
  \bibinfo{journal}{D-Lib Magazine} \textbf{\bibinfo{volume}{10}}
  (\bibinfo{year}{2004}).

\bibitem[{\citenamefont{Eysenbach}(2006)}]{eysenbach-2006}
\bibinfo{author}{\bibfnamefont{G.}~\bibnamefont{Eysenbach}},
  \bibinfo{journal}{PLoS Biol} \textbf{\bibinfo{volume}{4(5)}}
  (\bibinfo{year}{2006}), \bibinfo{note}{e157}.

\bibitem[{\citenamefont{von Neumann and
  Morgenstern}(1947)}]{book_neumann_gametheory}
\bibinfo{author}{\bibfnamefont{J.}~\bibnamefont{von Neumann}} \bibnamefont{and}
  \bibinfo{author}{\bibfnamefont{O.}~\bibnamefont{Morgenstern}},
  \emph{\bibinfo{title}{The Theory of Games and Economic Behaviour}}
  (\bibinfo{publisher}{Princeton University Press}, \bibinfo{year}{1947}).

\bibitem[{\citenamefont{von Neumann}(1932)}]{book_neumann_quantumtheory}
\bibinfo{author}{\bibfnamefont{J.}~\bibnamefont{von Neumann}},
  \emph{\bibinfo{title}{Mathematische Grundlagen der Quantenmechanik}}
  (\bibinfo{publisher}{Springer}, \bibinfo{year}{1932}).

\bibitem[{\citenamefont{Meyer}(1999)}]{meyer-1999-82}
\bibinfo{author}{\bibfnamefont{D.~A.} \bibnamefont{Meyer}},
  \bibinfo{journal}{Physical Review Letters} \textbf{\bibinfo{volume}{82}},
  \bibinfo{pages}{1052} (\bibinfo{year}{1999}), \eprint{\href{http://xxx.lanl.gov/abs/quant-ph/9804010}{quant-ph/9804010}}.

\bibitem[{\citenamefont{Eisert et~al.}(1999)\citenamefont{Eisert, Wilkens, and
  Lewenstein}}]{eisert-1999-83}
\bibinfo{author}{\bibfnamefont{J.}~\bibnamefont{Eisert}},
  \bibinfo{author}{\bibfnamefont{M.}~\bibnamefont{Wilkens}}, \bibnamefont{and}
  \bibinfo{author}{\bibfnamefont{M.}~\bibnamefont{Lewenstein}},
  \bibinfo{journal}{Physical Review Letters} \textbf{\bibinfo{volume}{83}},
  \bibinfo{pages}{3077} (\bibinfo{year}{1999}), \eprint{\href{http://xxx.lanl.gov/abs/quant-ph/9806088}{quant-ph/9806088}}.

\bibitem[{\citenamefont{Benjamin and Hayden}(2001)}]{benjamin-2001-64}
\bibinfo{author}{\bibfnamefont{S.~C.} \bibnamefont{Benjamin}} \bibnamefont{and}
  \bibinfo{author}{\bibfnamefont{P.~M.} \bibnamefont{Hayden}},
  \bibinfo{journal}{Physical Review A} \textbf{\bibinfo{volume}{64}},
  \bibinfo{pages}{030301} (\bibinfo{year}{2001}), \eprint{\href{http://xxx.lanl.gov/abs/quant-ph/0007038}{quant-ph/0007038}}.

\bibitem[{\citenamefont{Marinatto and Weber}(2000)}]{marinatto-2000-272}
\bibinfo{author}{\bibfnamefont{L.}~\bibnamefont{Marinatto}} \bibnamefont{and}
  \bibinfo{author}{\bibfnamefont{T.}~\bibnamefont{Weber}},
  \bibinfo{journal}{Physics Letters A} \textbf{\bibinfo{volume}{272}},
  \bibinfo{pages}{291} (\bibinfo{year}{2000}), \eprint{\href{http://xxx.lanl.gov/abs/quant-ph/0004081}{quant-ph/0004081}}.

\bibitem[{\citenamefont{Piotrowski and Sladkowski}(2002)}]{piotrowski-2002-312}
\bibinfo{author}{\bibfnamefont{E.~W.} \bibnamefont{Piotrowski}}
  \bibnamefont{and}
  \bibinfo{author}{\bibfnamefont{J.}~\bibnamefont{Sladkowski}},
  \bibinfo{journal}{Physica A} \textbf{\bibinfo{volume}{312}},
  \bibinfo{pages}{208} (\bibinfo{year}{2002}), \eprint{\href{http://xxx.lanl.gov/abs/quant-ph/0104006}{quant-ph/0104006}}.

\bibitem[{\citenamefont{Du et~al.}(2002)\citenamefont{Du, Li, Xu, Shi, Wu,
  Zhou, and Han}}]{du-2002-88}
\bibinfo{author}{\bibfnamefont{J.}~\bibnamefont{Du}},
  \bibinfo{author}{\bibfnamefont{H.}~\bibnamefont{Li}},
  \bibinfo{author}{\bibfnamefont{X.}~\bibnamefont{Xu}},
  \bibinfo{author}{\bibfnamefont{M.}~\bibnamefont{Shi}},
  \bibinfo{author}{\bibfnamefont{J.}~\bibnamefont{Wu}},
  \bibinfo{author}{\bibfnamefont{X.}~\bibnamefont{Zhou}}, \bibnamefont{and}
  \bibinfo{author}{\bibfnamefont{R.}~\bibnamefont{Han}},
  \bibinfo{journal}{Physical Review Letters} \textbf{\bibinfo{volume}{88}},
  \bibinfo{pages}{137902} (\bibinfo{year}{2002}), \eprint{\href{http://xxx.lanl.gov/abs/quant-ph/0104087}{quant-ph/0104087}}.

\bibitem[{\citenamefont{Iqbal}(2006)}]{iqbal-2006}
\bibinfo{author}{\bibfnamefont{A.}~\bibnamefont{Iqbal}}, Ph.D. thesis,
  \bibinfo{school}{University of Hull (UK)} (\bibinfo{year}{2006}),
  \eprint{\href{http://xxx.lanl.gov/abs/quant-ph/0604188}{quant-ph/0604188}}.

\bibitem[{\citenamefont{Grabbe}(2005)}]{grabbe-2005}
\bibinfo{author}{\bibfnamefont{J.~O.} \bibnamefont{Grabbe}},
  \emph{\bibinfo{title}{An introduction to quantum game theory}}
  (\bibinfo{year}{2005}), \eprint{\href{http://xxx.lanl.gov/abs/quant-ph/0506219}{quant-ph/0506219}}.

\bibitem[{\citenamefont{Flitney and Abbott}(2002)}]{flitney-2002-2}
\bibinfo{author}{\bibfnamefont{A.~P.} \bibnamefont{Flitney}} \bibnamefont{and}
  \bibinfo{author}{\bibfnamefont{D.}~\bibnamefont{Abbott}},
  \bibinfo{journal}{Fluct. Noise Lett.} \textbf{\bibinfo{volume}{2}},
  \bibinfo{pages}{R175} (\bibinfo{year}{2002}), \eprint{\href{http://xxx.lanl.gov/abs/quant-ph/0208069}{quant-ph/0208069}}.

\bibitem[{\citenamefont{Eisert and Wilkens}(2000)}]{eisert-2000-47}
\bibinfo{author}{\bibfnamefont{J.}~\bibnamefont{Eisert}} \bibnamefont{and}
  \bibinfo{author}{\bibfnamefont{M.}~\bibnamefont{Wilkens}},
  \bibinfo{journal}{Journal of Modern Optics} \textbf{\bibinfo{volume}{47}},
  \bibinfo{pages}{2543} (\bibinfo{year}{2000}), \eprint{\href{http://xxx.lanl.gov/abs/quant-ph/0004076}{quant-ph/0004076}}.

\end{thebibliography}
\end{document}